\documentclass[12pt]{article}
\usepackage{amsmath}
\usepackage{times}
\usepackage{mathrsfs}
\usepackage{amssymb}
\usepackage{graphicx}
\usepackage{color}
\usepackage{multirow}
\usepackage[authoryear]{natbib}
\usepackage{rotating}
\usepackage{bbm}
\usepackage{latexsym}
\usepackage[unicode=true,
 bookmarks=true,bookmarksnumbered=false,bookmarksopen=false,
 breaklinks=false,pdfborder={0 0 1},backref=false,colorlinks=false]
 {hyperref}
\hypersetup{pdftitle={Spontaneous Motion on Two-dimensional Continuous Attractors}}
\newcommand{\mydot}{.}

\newcommand{\captionfonts}{\normalsize}

\allowdisplaybreaks

\makeatletter  
\long\def\@makecaption#1#2{%
  \vskip\abovecaptionskip
  \sbox\@tempboxa{{\captionfonts #1: #2}}%
  \ifdim \wd\@tempboxa >\hsize
    {\captionfonts #1: #2\par}
  \else
    \hbox to\hsize{\hfil\box\@tempboxa\hfil}%
  \fi
  \vskip\belowcaptionskip}
\makeatother   

\newcommand{\etal}{{\it et~al.}}

\begin{document}
\hspace{13.9cm}1

\ \vspace{20mm}\\

{\bf Spontaneous Motion on Two-dimensional Continuous Attractors}

\ \\
{\bf C. C. Alan Fung$^{\displaystyle 1}$ and S. Amari$^{\displaystyle 2}$}\\
{$^{\displaystyle 1}$Department of Physics, The Hong Kong University
of Science and Technology, Clear Water Bay, Hong Kong, China.}\\
{$^{\displaystyle 2}$Laboratory for Mathematical Neuroscience, RIKEN
Brain Science Institute, Saitama, 351-0198, Japan.}\\
%


\thispagestyle{empty}
\markboth{}{NC instructions}
\ \vspace{-0mm}\\
%
\begin{center} {\bf Abstract} \end{center}
Attractor models are simplified models used to describe the dynamics of
firing rate profiles of a pool of neurons. The firing rate profile,
or the neuronal activity, is thought to carry information. Continuous
attractor neural networks (CANNs) describe the neural processing
of continuous information such as object position, object orientation
and direction of object motion. Recently, it was found 
that, in one-dimensional CANNs, short-term synaptic depression 
can destabilize bump-shaped neuronal attractor activity profiles. 
In this paper, we study two-dimensional CANNs with short-term 
synaptic depression and with spike frequency adaptation.  
We found that the dynamics of CANNs with short-term synaptic depression 
and CANNs with spike frequency adaptation are qualitatively similar. 
We also found that in both kinds of CANNs the perturbative approach
can be used to predict phase diagrams, dynamical variables and speed of 
spontaneous motion.

\section{Introduction}

Neurons communicate with each other through the neurotransmitter
diffusion initiated by action potentials, or \textit{spikes}, and
the activity of one neuron can excite or inhibit the activity of 
another neurons. The firing rates of spike trains are thought to 
carry information and correlations between the firing rates of 
neurons depend on the strength of the couplings between those neurons. 
With some specific settings of couplings (e.g., Mexican-hat
couplings \citep{Amari1977,Ben-Yishai1995}) networks of neurons can
support a continuous family of local neuronal activity profiles on a 
field, which can be used to represent continuous information, such as
object position, object orientation and direction of object motion.

Local neuronal activities associated with continuous information
are observed in various brain regions. Typical examples of cells 
showing such activity are head-direction cells 
\citep{Taube1990,Blair1995,Zhang1996,Taube1998}, place cells
\citep{OKeefe1971,OKeefe1976,OKeefe1996,Samsonovich1997} and 
moving-direction cells \citep{Maunsell1983,Treue2000}. 
These cells have Gaussian-like tuning curves as functions of stimulus. 
Among the numerous models proposed to describe behaviors of these
systems are continuous attractor neural networks (CANNs), and a recent 
study of persistent activity in monkey prefrontal cortices has provided
evidence of continuous attractors in the central nervous system.
\citep{Wimmer2014}. Persistent neuronal activity in monkeys' 
prefrontal cortices was discovered during delayed-response 
tasks \citep{Funahashi1993}.  The study by \citet{Wimmer2014}
confirmed that pairwise neuronal correlation predicted by
theories can be observed in the brain region they investigated 
\citep[][]{Ben-Yishai1995,Pouget1998,Wu2008}.
  
\begin{figure}
\begin{centering}
\includegraphics[width=1\textwidth]{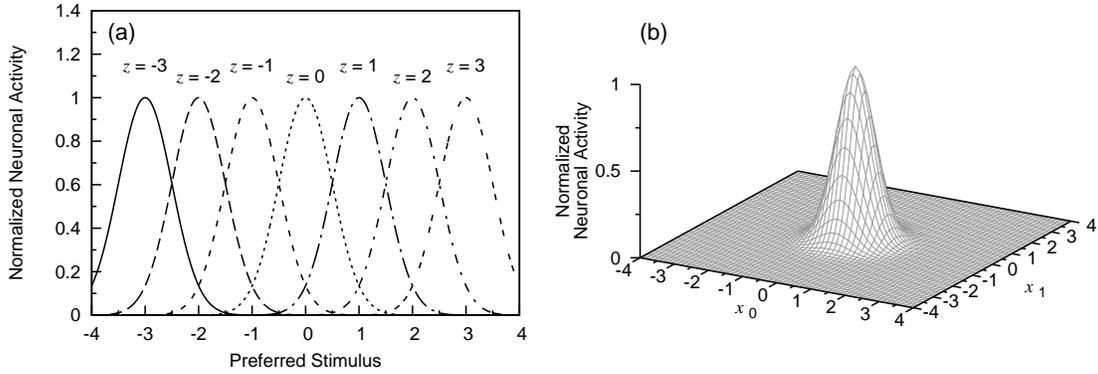}\protect\caption{\label{fig:canns} 
Examples of attractor states of continuous attractor
neural networks (CANNs) in (a) a 1D field and (b) a 2D field. 
In (a), profiles are centered at location $z$, the 
location they are representing. In (b), $x_{0}$ is the first 
coordinate of the preferred stimulus and $x_{1}$ is the second 
coordinate of the preferred stimulus.}

\par\end{centering}

\end{figure}

A one-dimensional (1D) CANN can support a family of Gaussian-like
neuronal activity profiles. They are attractors of the network. As
shown in Figure \ref{fig:canns}(a), attractors have the same shapes
and are centered at positions corresponding to different preferred stimuli.
These Gaussian-like profiles can shift smoothly along the space of
attractors. Families of attractor profiles in two-dimensional (2D) 
CANNs are similar: each attractor profile is a Gaussian-like profile 
and centered at a particular position in the 2D field 
(Figure \ref{fig:canns}(b)). The dynamics of a network state profile 
to track a stimulus is widely studied in the literature 
\citep{Ben-Yishai1995,Samsonovich1997,Wu2008,Fung2010}. If couplings
between neurons are static over time (quenched), the steady state of 
neuronal activity profiles will be static because of the homogeneity and
translational invariance of CANNs.  If they depend on the firing 
histories of presynaptic neurons, however, the dynamics of CANNs can 
be different.

\citet{Tsodyks1997} proposed a model in which the synaptic
efficacies between neurons depend on the amount of available 
neurotransmitters in the presynaptic neurons and this amount 
depends on the firing history of the presynaptic neuron 
\citep{Tsodyks1997,Tsodyks1998}. This kind of reduction
in synaptic efficacies, due to past presynaptic neuronal activity, is
called short-term synaptic depression (STD). There are reports that
short-term synaptic depression can exhibit rich dynamics in CANNs
\citep{York2009,Fung2012a}. \citet{Fung2012a} reported that in
1D CANNs, short-term synaptic depression can destabilize attractor
profiles. Within a broad range of strengths of divisive
global inhibition, if we increase the degree of STD to a moderate
range, static activity profiles will be translationally destabilized. 
After some translational perturbations, a bump-shaped profile
can move spontaneously along the attractor space. In these scenarios,
both bump-shaped static states and moving states can coexist.
If we further increase the degree of STD, no static profile can be
found. If the degree of STD is too large, the steady state of the 
system can only be a trivial solution. Similar behavior is reported 
by \citet{York2009} with a different model. We can see that the 
instability induced by STD can reshape the intrinsic dynamics 
of the system, even when no stimulus is presented.

Network response in a CANN with static couplings is always lagging 
behind a continuously moving stimulus.  However, with short-term 
synaptic depression, due to the translational instability, the 
underlying dynamics of the network can make the response to over-take
the actual stimulus. \citet{Fung2012b} suggested that this behavior
can be used to implement a delay compensation mechanism.  Short-term 
synaptic depression can also induce global \citep{Leobel2002} and local 
\citep{Fung2013,Wang2014} periodic excitements of neuronal activity 
profiles.  These periodic excitements can enhance information processing 
in the brain. \citet{Fung2013} recently proposed that periodic excitement 
driven by the intrinsic dynamics can improve the resolution of CANNs. 
\citet{Kilpatrick2012} also proposed that the neuronal activity pattern 
may shift between one stimulus and another. These theories suggested
that STD may enhance the capability of CANNs.

Studies on CANNs with short-term synaptic depression are mainly on
1D networks. In this paper, we discuss the intrinsic dynamics
of bump-shaped solutions of two-dimensional CANNs with short-term
synaptic depression. Other 2D models possess rich dynamical behaviors
such as spiral waves \citep{Kilpatrick2010}, breathing pulses 
\citep{Kilpatrick2010} and collisions of two bump-shaped profiles 
\citep{Lu2011}. Here, we focus on the spontaneous motion of a single 
bump-shaped profile, and analyze its stability.  We have also studied
the influence of STD on the sizes of bump-shaped profiles and on their
changes in shape.

We study not only CANNs with STD, but also CANNs with spike frequency
adaptation. Spike frequency adaptation (SFA) is a dynamical feature
commonly observed in neurons.  Neurons are suppressed after prolonged
firing.  SFA can be generated by a number of mechanisms 
\citep{Brown1980,Madison1984,Fleidervish1996,Benda2003}.
It can also destabilize the amplitudes and positions of static bumps. 
SFA-induced destabilization of bump-shaped states in CANNs are reported
in the literature \citep[e.g.,][]{Kilpatrick2010b}. 
What we find in our study on CANNs with SFA is similar to what we 
found in the case with STD.  SFA first destabilizes the translational 
mode and then the amplitudal mode. There is also a parameter region 
such that both spontaneous-moving-bump solutions and static-bump
solutions can coexist.

In this paper, in each case, CANNs with STD and CANNs with SFA, 
we first introduce the model we used to study the problem and then 
analyze each scenario using the perturbative method proposed by 
\cite{Fung2010}. These sections are followed by a section discussing 
the comparison between theoretical and simulation results and 
discussing the limitations of the perturbative method.

\section{The Model}

In this work we consider a 2D neural field, where neurons are located 
on a 2D field with positional coordinates $\mathbf{x}=\left(x_{0},x_{1}\right)$.
$\mathbf{x}$ can be interpreted as the preferred stimulus 
of the neuron sitting at that point in the 2D field.
The state of a neural field is specified by the average membrane potential 
$u\left(\mathbf{x},t\right)$ of neurons at $\mathbf{x}$ at time $t$.
The neuronal activity of a neuron at $\mathbf{x}$ is given by a nonlinear
function of $u\left(\mathbf{x},t\right)$:
\begin{equation}
r\left(\mathbf{x},t\right)=\frac{u\left(\mathbf{x},t\right)^{2}}{B\left(t\right)}\Theta\left[u\left(\mathbf{x},t\right)\right],\label{eq:r_of_u}
\end{equation}
where $\Theta$ is the Heaviside step function and $B\left(t\right)$ is 
the divisive global inhibition. The neuronal activity is related 
to the average firing rate of the neuron. The evolution of the 
global inhibition is given by
\begin{equation}
\tau_{B}\frac{dB\left(t\right)}{dt}=-B\left(t\right)+1+\rho k\int d\mathbf{x}^{\prime}u\left(\mathbf{x}^{\prime},t\right)^{2}\Theta\left[u\left(\mathbf{x}^\prime,t\right)\right],\label{dBdt}
\end{equation}
where $k$ is the parameter controlling the strength of the divisive global
inhibition and $\rho$ is the density of neurons over the field. 
Here, $B\left(t\right)$ is driven by $u\left(\mathbf{x},t\right)^2$.
This choice of driving term can simplify our calculations.
As we will describe in the following, $u\left(\mathbf{x},t\right)$
is a weighted sum of $r\left(\mathbf{x},t\right)$.  So $B\left(t\right)$
effectively depends on $r\left(\mathbf{x},t\right)$.  
Also, at the steady state, the magnitude of $u\left(\mathbf{x},t\right)$ 
is directly proportional to that of $r\left(\mathbf{x},t\right)$.

In the network, neurons are connected by excitatory couplings given by
\begin{equation}
J\left(\left|\mathbf{x}-\mathbf{x^{\prime}}\right|\right)=\frac{J_{0}}{2\pi a^{2}}\exp\left(-\frac{\left|\mathbf{x}-\mathbf{x^{\prime}}\right|^{2}}{2a^{2}}\right),\label{eq:Jxx}
\end{equation}
where $\left|~\cdot~\right|$ is the norm of the argument. $a$ is the 
radius of effective excitatory connections and $J_{0}$ is the
intensity of average coupling over the field. The dynamics of
$u\left(x,t\right)$ is governed by 
\begin{align}
\tau_{s}\frac{\partial u\left(\mathbf{x},t\right)}{\partial t}=&-u\left(\mathbf{x},t\right)+\rho\int d\mathbf{x^{\prime}}J\left(\left|\mathbf{x}-\mathbf{x^{\prime}}\right|\right)p\left(\mathbf{x^{\prime}},t\right)r\left(\mathbf{x^{\prime}},t\right)\nonumber \\&-v\left(\mathbf{x},t\right)+I^{\text{ext}}\left(\mathbf{x},t\right). \label{eq:dudt}
\end{align}
$I^{\text{ext}}\left(\mathbf{x},t\right)$ is the external 
stimulus, $p\left(\mathbf{x},t\right)$ is the portion 
of available neurotransmitters of the presynaptic neuron and 
$v\left(\mathbf{x},t\right)$ is the dynamical variable 
corresponding to SFA.  Since in this work, we are studying 
intrinsic dynamics in the network, $I^{\rm ext}\left(\mathbf{x},t\right)$
is set to zero throughout the paper.

The portion of available neurotransmitters of 
the presynaptic neuron at $\mathbf{x}$, $p\left(\mathbf{x},t\right)$,
evolves as 
\begin{equation}
\tau_{d}\frac{\partial p\left(\mathbf{x},t\right)}{\partial t}=-p\left(\mathbf{x},t\right)+1-\tau_{d}\beta p\left(\mathbf{x},t\right)r\left(\mathbf{x},t\right).\label{eq:dpdt}
\end{equation}
The first term on the right-hand side of this equation is the relaxation of 
$p\left(\mathbf{x},t\right)$ with time constant $\tau_{d}$. 
The last term is the consumption rate of neurotransmitters. 
$\beta$ is a parameter proportional to the neurotransmitter
consumption due to each spikes. This parameter controls
the strength of STD.  On the other hand, the dynamics 
of $v\left(\mathbf{x},t\right)$ is given by
\begin{equation}
\tau_{i}\frac{dv}{dt}\left(\mathbf{x},t\right)=-v\left(\mathbf{x},t\right)+\gamma f\left[u\left(\mathbf{x},t\right)\right].\label{eq:dvdt_sub}
\end{equation}
$\tau_{i}$ is the timescale of SFA. $\gamma$ is the degree of SFA. 
$f$ is the dependence of SFA variables on average membrane potential,
which is a non-decreasing function. For simplicity, we have chosen
\begin{equation}
f\left(u\right)=u\Theta\left(u\right).\label{eq:f_u2v}
\end{equation}
This choice is convenient for our analytic purpose and should not 
affect the main conclusion qualitatively as long as the adaptation 
increases with the average membrane potential, $u\left(\mathbf{x},t\right)$, 
which is correlated with the average neuronal activity.

In the present study, we analyze CANNs with STD and CANNs with SFA 
separately. For the case of CANNs with STD, we set $\gamma=0$. For
CANNs with SFA, we set $\beta=0$.

\section{CANNs with STD}
\subsection{Stationary Solution}

\begin{figure}
\begin{centering}
\includegraphics[width=0.8\textwidth]{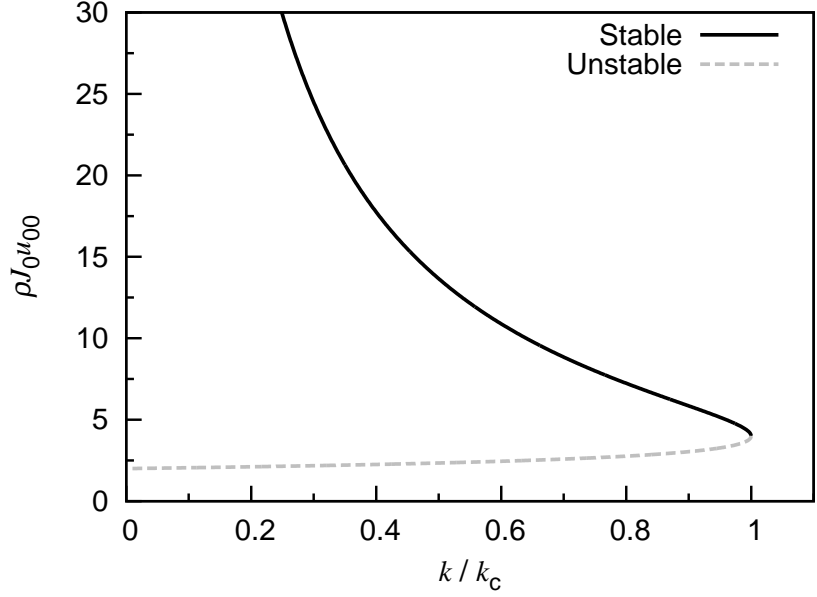}
\par\end{centering}

\protect\caption{\label{fig:static_no_std} Solution of $\rho J_{0}u_{00}$ as a function
of $k/k_{c}$. For each $k/k_{c}$ less than 1, the larger solution of
$\rho J_{0}u_{00}$ is stable and the smaller solution is unstable.}
\end{figure}

In this case, we set $\gamma=0$ and $I^{\rm ext}\left(\mathbf{x},t\right)=0$ 
to suppress SFA at the moment. For $\beta=0$, $p\left(\mathbf{x},t\right)=1$,
two types of non-zero fixed point solutions to Eq. (\ref{eq:dudt})
exist when $0<k<k_{c}\equiv\rho J_{0}^{2}/\left(32\pi a\right)$:
\begin{equation}
u\left(\mathbf{x}\right)=u_{00}\exp\left(-\frac{\left|\mathbf{x}-\mathbf{z}\right|^{2}}{4a^{2}}\right),
\end{equation}
where $\rho J_{0}u_{00}=4(1\pm\sqrt{1-k/k_{c}})/(k/k_{c})$ and $\mathbf{z}$
is an arbitrary location in the field representing the center of local
excitation. It is expected that the fixed point solution with the
larger amplitude is stable and the other is unstable. For $\tau_{B}=0$,
consider 
\begin{equation}
u\left(\mathbf{x},t\right)=\left[u_{00}+\delta u_{00}\left(t\right)\right]\exp\left(-\frac{\left|\mathbf{x}-\mathbf{z}\right|^{2}}{4a^{2}}\right),\label{eq:u_plus_delta_u}
\end{equation}
where $\delta u_{00}\left(t\right)$ is the deviation of the profile
from the fixed point solution. Then the differential equation of 
the deviation is
\begin{equation}
\tau_{s}\frac{d\delta u_{00}\left(t\right)}{dt}=\mp\sqrt{1-\frac{k}{k_{c}}}\delta u_{00}\left(t\right).
\end{equation}
Therefore the smaller fixed point solutions are unstable, while the
larger fixed point solutions have stable amplitudes. For non-zero
$\tau_{B}$, we have to consider the dynamics of $B(t)$. Let $B(t)=B_{0}+\delta B(t)$,
where $B_{0}=1+\tfrac{1}{16}\left(k/k_{c}\right)\left(\rho J_{0}u_{00}\right)^{2}$.
The dynamics of the deviation from the fixed point solution is given
by 
\begin{equation}
\tau_{s}\frac{d}{dt}\left(\begin{array}{c}
\rho J_{0}\delta u_{00}\left(t\right)\\
\delta B\left(t\right)
\end{array}\right)=\left(\begin{array}{cc}
1 & -2\\
\frac{1}{2}\frac{\tau_{s}}{\tau_{B}}(1\pm\sqrt{1-\frac{k}{k_{c}}}) & -\frac{\tau_{s}}{\tau_{B}}
\end{array}\right)\left(\begin{array}{c}
\rho J_{0}\delta u_{00}\left(t\right)\\
\delta B\left(t\right)
\end{array}\right).\label{eq:ddelta_u}
\end{equation}
Since the trace of the $2\times2$ matrix in Eq. (\ref{eq:ddelta_u}) is negative
for $\tau_{B}<\tau_{s}$ and the sign of the determinant is independent
of $\tau_{s}/\tau_{B}$, the solution with the larger $u_{00}$ is
stable against perturbations in amplitude, while the solution with the
smaller amplitude is unstable. The solution of $u_{00}$ as a function
of $k/k_{c}$ is shown in Figure \ref{fig:static_no_std}.

For $\beta>0$ and $I^{\text{ext}}\left(\mathbf{x},t\right)=0$, we
approximate the attractor profiles of $u\left(\mathbf{x},t\right)$
and $1-p(\mathbf{x},t)$ by non-moving Gaussian distributions
\begin{align}
u\left(\mathbf{x},t\right)  = &~u_{00}\left(t\right)\exp\left(-\frac{\left|\mathbf{x}\right|^{2}}{4a^{2}}\right),\label{eq:u0}\\
1-p\left(\mathbf{x},t\right)  = &~p_{00}\left(t\right)\exp\left(-\frac{\left|\mathbf{x}\right|^{2}}{2a^{2}}\right).\label{eq:p0}
\end{align}
Here, without loss of generality, we consider the case with 
$\mathbf{z}=\mathbf{0}$. They are not the exact solutions of Eqs. 
(\ref{dBdt}), (\ref{eq:dudt}) and (\ref{eq:dpdt}). In this
ansatz, $u\left(\mathbf{x},t\right)$ is assumed to have the same 
shape as that in the $\beta=0$ case. The width of 
$1-p\left(\mathbf{x},t\right)$ is different from 
$u\left(\mathbf{x},t\right)$ because the shape of 
$1-p\left(\mathbf{x},t\right)$ is similar to $u\left(\mathbf{x},t\right)^2$ at the 
small $\beta$ limit. The differential equations governing $u_{00}\left(t\right)$ 
and $p_{00}\left(t\right)$ can be obtained by projecting Eq. (\ref{eq:dudt}) onto 
$\exp[-\left|\mathbf{x}\right|^{2}/(4a^{2})]$ and projecting Eq. (\ref{eq:dpdt})
onto $\exp[-\left|\mathbf{x}\right|^{2}/(2a^{2})]$.

As in the study by \citet{Fung2012a}, $u\left(\mathbf{x},t\right)$ 
can be replaced by rescaled variables $\widetilde{u}\left(\mathbf{x},t\right)\equiv
\rho J_0 u\left(\mathbf{x},t\right)$ because $u_{00}\left(t\right)$
has a dimension $1/(\rho J_{0})$. $\widetilde{u}_{00}\left(t\right)\equiv\rho J_{0}u_{00}\left(t\right)$.
And $k$ and $\beta$ can be rescaled by $\widetilde{k}\equiv k/k_{c}$
and $\widetilde{\beta}\equiv\tau_{d}\beta/(\rho^{2}J_{0}^{2})$.

\begin{figure}
\begin{centering}
\includegraphics[width=0.8\columnwidth]{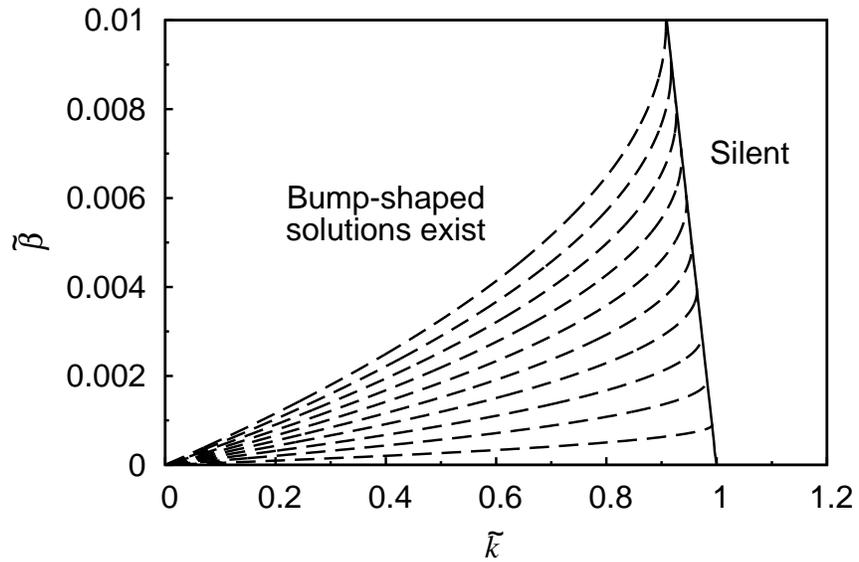}
\par\end{centering}

\protect\caption{\label{fig:std_parabola} Parameter regions of existence of static
profiles of $u(\mathbf{x},t)$. Dashed curves: parabolas given by
Eq. (\ref{eq:std_parabola}) for different $\widetilde{\beta}B$'s. Solid
line: boundary separating parameter regions of static profiles of $u(\mathbf{x},t)$
and silent phase. }

\end{figure}

With these rescaled variables, the differential equations of 
$\widetilde{u}_{00}\left(t\right)$, $p_{00}\left(t\right)$ and 
$B\left(t\right)$ are
\begin{align}
\tau_{s}\frac{d\widetilde{u}_{00}}{dt}\left(t\right)  = & ~ -\widetilde{u}_{00}\left(t\right)+\frac{1}{2}\frac{1}{B\left(t\right)}\widetilde{u}_{00}\left(t\right)^{2}\left[1-\frac{4}{7}p_{00}\left(t\right)\right],\label{eq:du00dt_std}\\
\tau_{d}\frac{dp_{00}}{dt}\left(t\right)  = & ~ -p_{00}\left(t\right)+\frac{1}{B\left(t\right)}\widetilde{\beta}\widetilde{u}_{00}\left(t\right)^{2}\left[1-\frac{2}{3}p_{00}\left(t\right)\right],\label{eq:dp00dt_std}\\
\tau_{B}\frac{dB}{dt}\left(t\right)  = & ~ -B\left(t\right)+1+\frac{1}{16}\widetilde{k}\widetilde{u}_{00}\left(t\right)^{2}.\label{eq:dBdt_std}
\end{align}
As we study the steady state behavior of the network, 
we need to solve for the fixed point solution first.
For the first two differential equations, we may let $\widetilde{\beta}B$
be given. Then we can solve $\widetilde{u}_{00}/B$ and $p_{00}$. For
a given $\widetilde{\beta}B$, $\widetilde{k}$ and $\widetilde{\beta}$ 
are related by
\begin{equation}
\widetilde{k}=\frac{16}{\left(\frac{\widetilde{u}_{00}}{B}\right)^{2}}\left[-\frac{1}{\left(B\widetilde{\beta}\right)^{2}}\widetilde{\beta}^{2}+\frac{1}{B\widetilde{\beta}}\widetilde{\beta}\right].\label{eq:std_parabola}
\end{equation}
The parabolas given by this equation for different 
$\widetilde\beta B$s are plotted as dashed lines
in Figure \ref{fig:std_parabola}, where we see that the static 
fixed point solution exists only when 
\begin{equation}
\begin{cases}
\widetilde{\beta}\le\frac{3}{16}\widetilde{k}\frac{1-\sqrt{\widetilde{k}}}{\sqrt{\widetilde{k}}-\frac{1}{7}} & \text{, if }\frac{9}{49}\le\widetilde{k}\le1\\
\widetilde{\beta}\le\frac{9}{56}\left(\frac{3}{7}-\sqrt{\frac{9}{49}-\widetilde{k}}\right) & \text{, if }0\le\widetilde{k}\le\frac{9}{49}
\end{cases}.
\end{equation}
By considering the stability of fixed point solutions, we obtain
the solid line in Figure \ref{fig:std_parabola} (also the dotted line 
in Figure \ref{fig:phase_diagram_std}).
This curve maps the parameter region for the existence of static 
profiles of $u(\mathbf{x},t)$. The methodology for studying 
stability of fixed point solutions can be found in 
Appendix \ref{subsec:static_std}.

\subsubsection{Translational Instability}

We have simplified Eqs. (\ref{dBdt}), (\ref{eq:dudt}) and (\ref{eq:dpdt})
by introducing the approximation given by Eqs. (\ref{eq:u0}) and
(\ref{eq:p0}). This simplification, however, is useful for studying 
only the amplitudal stability of a bump-shaped solution. For the
translational stability, we need to consider the stability of the static
solution against asymmetric distortions.
Displacing originally aligned $u\left(\mathbf{x},t\right)$ and
$p\left(\mathbf{x},t\right)$ profiles of a static solution is
a reasonable test, as the dip of $p\left(\mathbf{x},t\right)$
is generated by activities of neurons. If the solution is moving,
the dip of $p\left(\mathbf{x},t\right)$ is always lagging behind.
As a result, the asymmetric component of $p\left(\mathbf{x},t\right)$
with respect to the center of mass of $u\left(\mathbf{x},t\right)$
becomes non-zero. In the calculation, we may drop asymmetric 
components of $u\left(\mathbf{x},t\right)$ for the moment, as we can always
choose a frame such that the major asymmetric mode is zero. 

Let us assume
\begin{align}
u\left(\mathbf{x},t\right)  = & ~ u_{00}\left(t\right)\exp\left(-\frac{\left|\mathbf{x}\right|^{2}}{4a^{2}}\right),\label{eq:u0_2}\\
p\left(\mathbf{x},t\right)  = & ~ p_{00}\left(t\right)\exp\left(-\frac{\left|\mathbf{x}\right|^{2}}{2a^{2}}\right)+p_{10}\left(t\right)\frac{x_{0}}{a}\exp\left(-\frac{\left|\mathbf{x}\right|^{2}}{2a^{2}}\right).\label{eq:p0p1}
\end{align}
Due to the symmetry of the preferred stimulus space, here we consider 
only the distortion along the $x_{0}$-direction. By substituting
these two assumptions into Eqs. (\ref{dBdt}), (\ref{eq:dudt}) and
(\ref{eq:dpdt}), we can study the stability of static solutions against 
asymmetric distortions. With assumptions Eqs. (\ref{eq:u0_2})
and (\ref{eq:p0p1}), we can derive the stability matrix by calculating
the Jacobian matrix at the fixed point solution. Then the dynamics
of distortions of $\widetilde u_{00}\left(t\right)$, 
$p_{00}\left(t\right)$, $p_{10}\left(t\right)$ and $B\left(t\right)$ 
around the static fixed point solution is given by
\begin{equation}
\tau_s
\frac{d}{dt}
\left(
\begin{array}{c}
\delta\widetilde u_{00}\left(t\right) \\
\delta p_{00}\left(t\right) \\
\delta B \left(t\right) \\
\delta p_{10}\left(t\right) 
\end{array}
\right)
=
\left(\begin{array}{cc}
\mathscr{A}_{\rm STD} & 0\\
0 & M
\end{array}\right)
\left(
\begin{array}{c}
\delta\widetilde u_{00}\left(t\right) \\
\delta p_{00}\left(t\right) \\
\delta B \left(t\right) \\
\delta p_{10}\left(t\right) 
\end{array}
\right),
\label{eq:trans_std}
\end{equation}
where the $3\times3$ matrix $\mathscr{A}_{\rm STD}$ is provided 
in Appendix \ref{subsec:static_std}.
$\mathscr{A}_{\rm STD}$ determines the amplitudal stability of 
the solution, while 
$M$ determines translational stability, which is given by
\begin{equation}
 M = \frac{1}{\tau_{d}}\left(-1+\frac{\widetilde{\beta}}{B}\frac{4}{9}\widetilde{u}_{00}^{2}+\frac{\tau_{d}}{\tau_{s}}\frac{1}{B}\frac{8}{49}\widetilde{u}_{00}\widetilde{p}_{00}\right).
\end{equation}
If $M>0$, the static solution of two-dimensional CANNs with short-term 
synaptic depression will obviously be translationally unstable because 
positional distortion will diverge.

\begin{figure}
\begin{centering}
\includegraphics[width=0.8\textwidth]{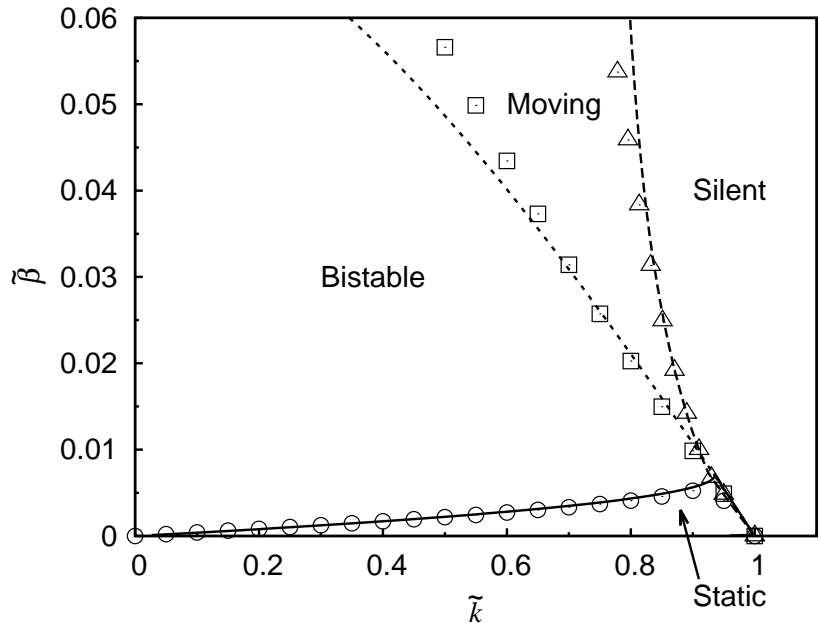}
\par\end{centering}

\begin{centering}
\protect\caption{\label{fig:phase_diagram_std}
Phase diagram of different phases under
this theoretical framework. 
Circles: simulation results on translational stability.
Squares: simulation results on amplitude stability.
Triangles: simulation results on stability of moving bumps.
Curves: corresponding theoretical predictions. 
Parameter: $\tau_{d}/\tau_{s}=50$.}
\end{centering}
\end{figure}

By studying the stability matrix, we found that bump-shaped solutions
will be translationally stable only if 
\begin{equation}
\frac{4}{9}\left(\widetilde{\beta}B\right)\left(\frac{\widetilde{u}_{00}}{B}\right)^{2}+\frac{\tau_{d}}{\tau_{s}}\frac{8}{49}\frac{\widetilde{u}_{00}}{B}p_{00}<1.
\end{equation}
In this case, asymmetric distortions cannot initiate spontaneous motion.
When this inequality does not hold, there will be a moving
solution such that the bump-shaped profile can move spontaneously
with a speed dictated by $\widetilde{k}$, $\widetilde{\beta}$ and 
$\tau_{d}/\tau_{s}$. This inequality provides a prediction of the 
boundary separating translationally stable static bumps and translationally
unstable static bumps, which is plotted as a solid line in 
Figure \ref{fig:phase_diagram_std}.

\begin{figure}
\begin{centering}
\includegraphics[width=\textwidth]{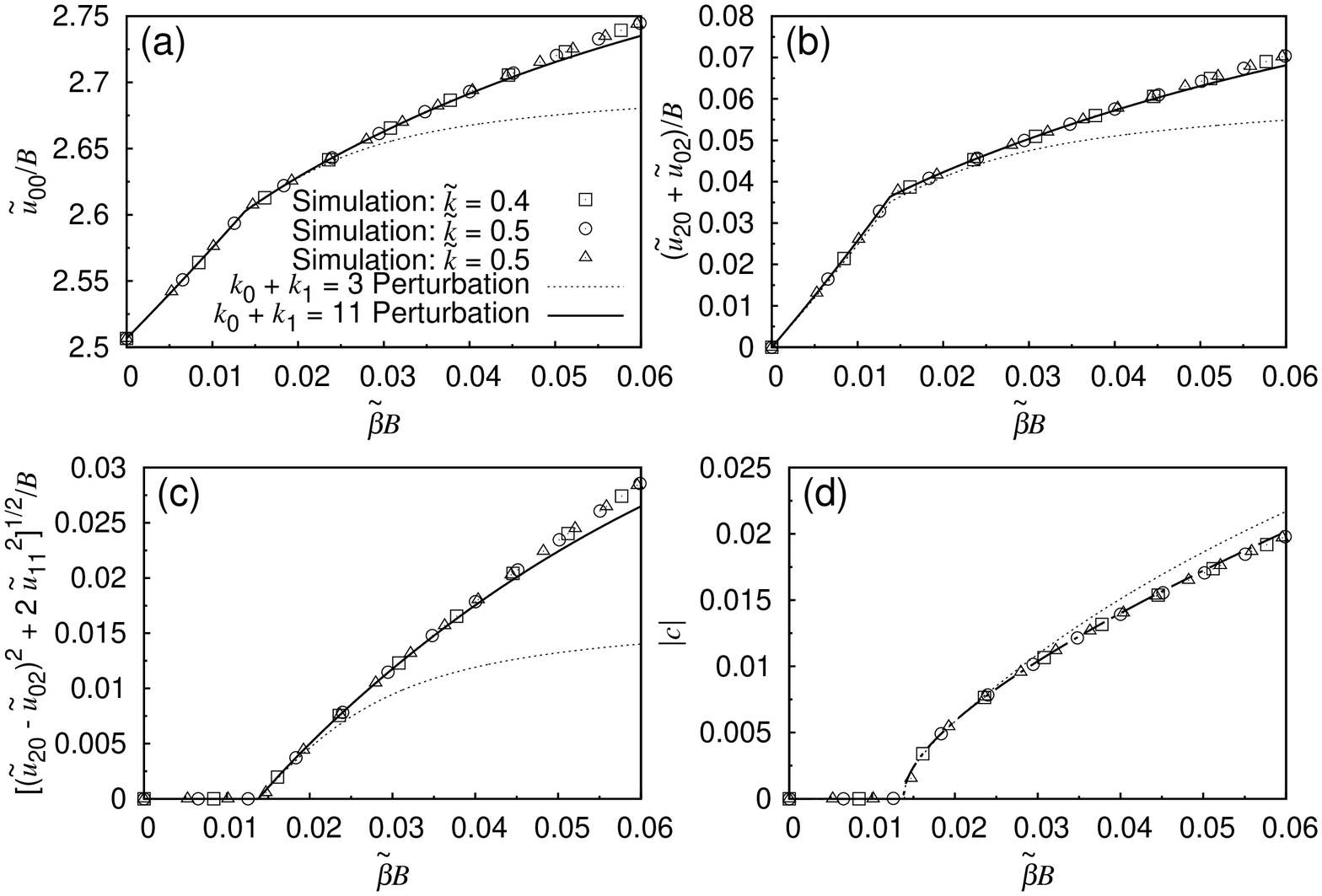}\protect
\caption{\label{fig:std_u00}
(a) Measurements and predictions of $\widetilde{u}_{00}/B$.
Symbols: measurements from various $\widetilde{k}$. Curves: different
levels of predictions. 
(b) Measurements and predictions of 
$\left(\widetilde{u}_{20}+\widetilde{u}_{02}\right)/B$.
Symbols: measurements from various $\widetilde{k}$. Curves: different
levels of predictions. 
(c) Measurements and predictions of $\sqrt{\left(\widetilde{u}_{20}-\widetilde{u}_{02}\right)^{2}+\widetilde{u}_{11}^{2}}/B$.
Symbols: measurements from various $\widetilde{k}$. Curves: different
levels of predictions. 
(d) Measurements and predictions of the intrinsic moving
speed. Symbols: measurements from various $\widetilde{k}$. Curves: different
levels of predictions. 
(a) - (d) Parameters: same as Figure \ref{fig:phase_diagram_std}.
}
\par\end{centering}

\end{figure}

\subsection{Moving Solution}

In the one-dimensional situation, when $\widetilde{k}$ is small and $\widetilde{\beta}$
is relatively large, there may be spontaneous moving solutions \citep{Fung2012a}.
To analyze the moving solution, we need to consider higher-order
expansions of $u\left(\mathbf{x},t\right)$ and $p\left(\mathbf{x},t\right)$.
In general, $u\left(\mathbf{x},t\right)$ and $p\left(\mathbf{x},t\right)$
can be expanded by any basis functions. Here we have chosen eigenstates
of quantum harmonic oscillator as basis functions.
\begin{align}
u\left(\mathbf{x},t\right)  = & ~ \sum_{k_{0},k_{1}}u_{k_{0}k_{1}}\left(t\right)\psi_{k_{0}}\left(\xi_{0}\right)\psi_{k_{1}}\left(\xi_{1}\right),\label{eq:uk}\\
1-p\left(\mathbf{x},t\right)  = & ~ \sum_{k_{0},k_{1}}p_{k_{0}k_{1}}\left(t\right)\varphi_{k_{0}}\left(\xi_{0}\right)\varphi_{k_{1}}\left(\xi_{1}\right),\label{eq:pk}
\end{align}
where 
\begin{align}
\psi_{k}\left(\xi_{i}\right)  = & ~ \frac{1}{\sqrt{\sqrt{2\pi}a2^{k}k!}}H_{n}\left(\frac{\xi_{i}}{\sqrt{2}a}\right)\exp\left[-\frac{\xi_{i}^{2}}{4a^{2}}\right],\label{eq:psi_k}\\
\varphi_{k}\left(\xi_{i}\right)  = & ~ \frac{1}{\sqrt{\sqrt{\pi}a2^{k}k!}}H_{n}\left(\frac{\xi_{i}}{a}\right)\exp\left[-\frac{\xi_{i}^{2}}{2a^{2}}\right],\; i=0,1.\label{eq:varphi_k}
\end{align}
In these equations, 
$\xi_{i}\equiv x_{i}-c_{i}t$. $c_{i}$ is the $i^{\rm th}$-component
of the velocity of the moving frame. $H_{n}$ is the $n^{\text{th}}$
order physicists' Hermite polynomial. As in the
ansatz for static solutions (Eqs. (\ref{eq:u0}) and (\ref{eq:p0})), 
the widths of the basis functions for ${u}\left(\mathbf{x},t\right)$
and $p\left(\mathbf{x},t\right)$ are different.  
This choice makes the perturbative expansion more efficient, 
although the choice of basis functions can be arbitrary.

After substituting Eqs. (\ref{eq:uk})
and (\ref{eq:pk}) into Eqs. (\ref{dBdt}), (\ref{eq:dudt}) 
and (\ref{eq:dpdt}) and projections on 
Eqs. (\ref{eq:psi_k}) and (\ref{eq:varphi_k}), we obtain
\begin{align}
 \tau_{s}\frac{d\widetilde{u}_{k_{0}k_{1}}}{dt}\left(t\right)  = & ~ -\widetilde{u}_{k_{0}k_{1}}\left(t\right)+\frac{\tau_{s}c_{0}}{2a}\left[\sqrt{k_{0}+1}\widetilde{u}_{k_{0}+1,k_{1}}\left(t\right)-\sqrt{k_{0}}\widetilde{u}_{k_{0}-1,k_{1}}\left(t\right)\right]\nonumber \\
 & ~  +\frac{\tau_{s}c_{1}}{2a}\left[\sqrt{k_{1}+1}\widetilde{u}_{k_{0},k_{1}+1}\left(t\right)-\sqrt{k_{1}}\widetilde{u}_{k_{0},k_{1}-1}\left(t\right)\right]\nonumber \\
 & ~  +\frac{1}{B\left(t\right)}\sum_{n_{0}n_{1}m_{0}m_{1}}C_{n_{0}m_{0}}^{k_{0}}C_{n_{1}m_{1}}^{k_{1}}\widetilde{u}_{n_{0}n_{1}}\left(t\right)\widetilde{u}_{m_{0}m_{1}}\left(t\right)\nonumber \\
& ~  -\frac{1}{B\left(t\right)}\sum_{n_{0}n_{1}m_{0}m_{1}l_{0}l_{1}}D_{n_{0}m_{0}l_{0}}^{k_{0}}D_{n_{1}m_{1}l_{1}}^{k_{1}}\widetilde{u}_{n_{0}n_{1}}\left(t\right)\widetilde{u}_{m_{0}m_{1}}\left(t\right)p_{l_{0}l_{1}}\left(t\right),\nonumber\\\label{eq:dukkdt}\\
 \tau_{d}\frac{dp_{k_{0}k_{1}}}{dt}\left(t\right)  = & ~ -p_{k_{0}k_{1}}\left(t\right)+\frac{\tau_{d}c_{0}}{\sqrt{2}a}\left[\sqrt{k_{0}+1}p_{k_{0}+1,k_{1}}\left(t\right)-\sqrt{k_{0}}p_{k_{0}-1,k_{1}}\left(t\right)\right]\nonumber \\
 & ~   +\frac{\tau_{d}c_{1}}{\sqrt{2}a}\left[\sqrt{k_{1}+1}p_{k_{0},k_{1}+1}\left(t\right)-\sqrt{k_{1}}p_{k_{0},k_{1}-1}\left(t\right)\right]\nonumber \\
 &  ~ +\frac{\widetilde{\beta}}{B\left(t\right)}\sum_{n_{0}n_{1}m_{0}m_{1}}E_{n_{0}m_{0}}^{k_{0}}E_{n_{1}m_{1}}^{k_{1}}\widetilde{u}_{n_{0}n_{1}}\left(t\right)\widetilde{u}_{m_{0}m_{1}}\left(t\right)\nonumber \\
 &~   -\frac{\widetilde{\beta}}{B\left(t\right)}\sum_{n_{0}n_{1}m_{0}m_{1}}F_{n_{0}m_{0}l_{0}}^{k_{0}}F_{n_{1}m_{1}l_{1}}^{k_{1}}\widetilde{u}_{n_{0}n_{1}}\left(t\right)\widetilde{u}_{m_{0}m_{1}}\left(t\right)p_{l_{0}l_{1}}\left(t\right),\label{eq:dpkkdt}\\
 \tau_{B}\frac{dB}{dt}\left(t\right) =  & ~ -B\left(t\right)+1+k\rho\sum_{k_{0}k_{1}}\widetilde{u}_{k_{0}k_{1}}\left(t\right)^{2}.\label{eq:dBdt}
\end{align}
A detailed illustration of the derivation of these equations can be found 
in Appendix \ref{subsec:moving_sol_std}.
Here, $\widetilde{u}_{k_{0}k_{1}}\left(t\right)$'s are rescaled dynamical variables:
$\widetilde{u}_{k_{0}k_{1}}\left(t\right)\equiv\rho J_{0}u_{k_{0}k_{1}}\left(t\right)$.
Coefficients
$C_{nm}^{k}$, $D_{nml}^{k}$, $E_{nm}^{k}$ and $F_{nml}^{k}$ are
defined by
\begin{align}
C_{nm}^{k}  \equiv & ~ \int d\xi\psi_{k}\left(\xi\right)\int d\xi^{\prime}\frac{1}{\sqrt{2\pi}a}e^{-\frac{\left(\xi-\xi^{\prime}\right)^{2}}{2a^{2}}}\psi_{n}\left(\xi^{\prime}\right)\psi_{m}\left(\xi^{\prime}\right),\label{eq:std_Cnmk}\\
D_{nml}^{k}  \equiv &~ \int d\xi\psi_{k}\left(\xi\right)\int d\xi^{\prime}\frac{1}{\sqrt{2\pi}a}e^{-\frac{\left(\xi-\xi^{\prime}\right)^{2}}{2a^{2}}}\psi_{n}\left(\xi^{\prime}\right)\psi_{m}\left(\xi^{\prime}\right)\varphi_{l}\left(\xi^{\prime}\right),\\
E_{nm}^{k}  \equiv &~ \int d\xi\varphi_{k}\left(\xi\right)\int d\xi^{\prime}\psi_{n}\left(\xi^{\prime}\right)\psi_{m}\left(\xi^{\prime}\right),\\
F_{nml}^{k}  \equiv &~ \int d\xi\varphi_{k}\left(\xi\right)\int d\xi^{\prime}\psi_{n}\left(\xi^{\prime}\right)\psi_{m}\left(\xi^{\prime}\right)\varphi_{l}\left(\xi^{\prime}\right).
\end{align}
$C_{00}^{0}$, $D_{000}^{0}$, $E_{00}^{0}$ and $F_{000}^{0}$ can
be calculated explicitly. For 
$C_{nm}^{k}$, $D_{nml}^{k}$, $E_{nm}^{k}$ and $F_{nml}^{k}$
with arbitrary $k$, $n$, $m$ and $l$, the
recurrence relations used to generate these coefficients can be found in
Appendix B in the paper by \citet{Fung2012a}. 
In practice, we cannot consider infinitely many 
$\widetilde{u}_{k_0k_1}\left(t\right)$ and $p_{k_0k_1}\left(t\right)$.  We 
used finite terms to obtain fairly acceptable results. In Figures
\ref{fig:phase_diagram_std} and \ref{fig:std_u00},
`$k_0+k_1=3$ Perturbation' means that we consider terms up to
$k_0+k_1=3$. 

Unfortunately, the above equations cannot
form a complete set of equations sufficient to solve for the fixed point
solution. We also need to consider the self-consistent condition:
\begin{equation}
\frac{\int d\boldsymbol{\xi}\widetilde{u}\left(\boldsymbol{\xi},t\right)\xi_{i}}{\int d\boldsymbol{\xi}\widetilde{u}\left(\boldsymbol{\xi},t\right)}=0.\label{eq:self-con}
\end{equation}
This self-consistent condition helps us to choose the center of mass
of $\widetilde{u}\left(\mathbf{x},t\right)$ to be origin of basis functions
so that the perturbative method can be efficient. 
$\widetilde{u}_{k_0k_1}/B$, $p_{k_0k_1}$ and $c_i$ can be solved
numerically by regarding $\widetilde{\beta}B$ as a constant and setting
time derivatives equal to zero.  
For each $\widetilde{\beta}B$, $\widetilde k$
and $\widetilde \beta$ are related by
\begin{equation}
\tilde{k}  = \frac{32\pi a^{2}}{\sum_{k_{0}k_{1}}\left(\frac{\widetilde{u}_{k_{0}k_{1}}\left(t\right)}{B}\right)^{2}}\left[\frac{1}{\widetilde{\beta}B}\widetilde{\beta}-\frac{1}{\left(\widetilde{\beta}B\right)^{2}}\widetilde{\beta}^{2}\right].
\end{equation}
Then fixed point solutions corresponding to each
$\left(\widetilde k,\widetilde\beta\right)$ can be solved.

\begin{figure}
\begin{centering}
\includegraphics[width=1\textwidth]{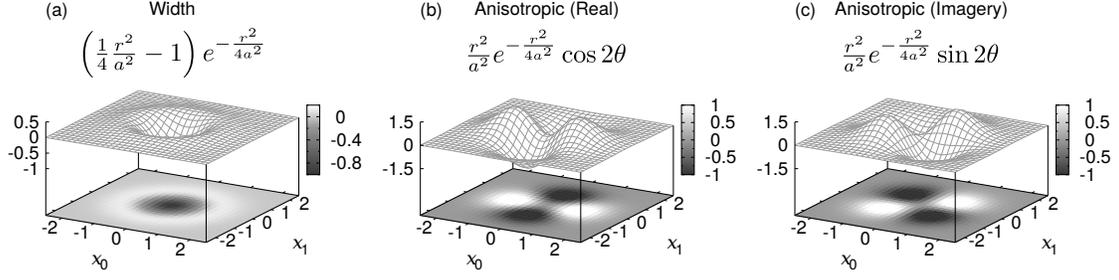}
\par\end{centering}

\protect\caption{\label{fig:std_basis_polar} Distortional modes for $k_{0}+k_{1}=2$.}
\end{figure}

By comparing the fixed point solution with the
simulation results, we found that not only fixed point solutions
but also various measurements of the network
state are determined by $\widetilde{\beta}B$ as shown in Figure \ref{fig:std_u00}. 
In Figures \ref{fig:std_u00}(b) and \ref{fig:std_u00}(c),
we compare the values of the second-order variables between simulations
and theoretical predictions by transforming them to the polar coordinates
via $\xi_{0}=r\cos\theta$ and $\xi_{1}=r\sin\theta$.
We can convert second-order basis functions for $\widetilde{u}\left(\mathbf{x},t\right)$
to be combinations of their polar counterparts. Then we can obtain a 
linear combination of basis functions in polar coordinates:
\begin{align}
   & ~ \widetilde{u}_{20}\psi_{2}\left(\xi_{0}\right)\psi_{0}\left(\xi_{1}\right)+\widetilde{u}_{02}\psi_{0}\left(\xi_{0}\right)\psi_{2}\left(\xi_{1}\right)+\widetilde{u}_{11}\psi_{1}\left(\xi_{0}\right)\psi_{1}\left(\xi_{1}\right)\nonumber \\
  = & ~ \left(\widetilde{u}_{20}+\widetilde{u}_{02}\right)\frac{1}{2a\sqrt{\pi}}\left(\frac{1}{4}\frac{r^{2}}{a^{2}}-1\right)e^{-\frac{r^{2}}{4a^{2}}}+\left(\widetilde{u}_{20}-\widetilde{u}_{02}-\sqrt{2}i\widetilde{u}_{11}\right)\frac{1}{8a\sqrt{\pi}}\frac{r^{2}}{a^{2}}e^{-\frac{r^{2}}{4a^{2}}}e^{i2\theta}\nonumber \\
   & ~ +\left(\widetilde{u}_{20}-\widetilde{u}_{02}+\sqrt{2}i\widetilde{u}_{11}\right)\frac{1}{8a\sqrt{\pi}}\frac{r^{2}}{a^{2}}e^{-\frac{r^{2}}{4a^{2}}}e^{-i2\theta}.\label{eq:std_rect2rot}
\end{align}
In order to compare predictions to simulation results without considering
the moving direction in the 2D field, we use $\left(\widetilde{u}_{20}+\widetilde{u}_{02}\right)/B$
(in Figure \ref{fig:std_u00}(b)) and \\
$\left|\widetilde{u}_{20}-\widetilde{u}_{02}-\sqrt{2}i\widetilde{u}_{11}\right|/B$
(in Figure \ref{fig:std_u00}(c)) for the case of $k_{0}+k_{1}=2$. 
$\left(\widetilde{u}_{20}+\widetilde{u}_{02}\right)$ is the
average change in the width of the bump-shaped profile relative to
the closest static solution, while $\left|\widetilde{u}_{20}-\widetilde{u}_{02}-\sqrt{2}i\widetilde{u}_{11}\right|$
is the magnitude of the anisotropic mode. Illustrations of the polar
functions in Eq. (\ref{eq:std_rect2rot}) are shown in Figure \ref{fig:std_basis_polar}.

Since $B$ is proportional to $\widetilde{u}_{00}^{2}$, an increase of
$\widetilde{u}_{00}/B$ implies a decrease in $\widetilde{u}_{00}$. In Figure
\ref{fig:std_u00}(a), the slope of $\widetilde{u}_{00}/B$ is discontinuous
at about $\widetilde{\beta}B\approx0.0137$. For $\widetilde{\beta}B\lesssim0.0137$,
the slope is significantly larger than that in the region of $\widetilde{\beta}B\gtrsim0.0137$.
This implies that the motion of a bump helps the bump to maintain its magnitude.
It also agrees with the tendency of the average membrane potential
(or neuronal activity) profile in 1D CANN that the bump tends to move to 
a region with a higher concentration of neurotransmitters \citep{York2009,Fung2012a}.

Figure \ref{fig:std_u00}(c) suggests that an anisotropic mode happens
only when the bump is not static, while Figure \ref{fig:std_u00}(b) shows
that the average width of $u(\mathbf{x},t)$ profile increases with the
strength of STD. The behavior of the average change in width, $\left(\widetilde{u}_{20}+\widetilde{u}_{02}\right)$,
is similar to that in height. The anisotropic mode happens only when
the bump is moving. This suggests that the bump get widened unevenly
due to the asymmetric $p(\mathbf{x},t)$ profile.

\subsection{Phase Diagram}

In the previous subsection, we have shown that the 
perturbative method can successfully predict different modes of 
distortions of $\widetilde{u}\left(\mathbf{x},t\right)$ and 
the intrinsic speed of spontaneous motion.  We have also
predicted the phase boundary separating translationally stable
static bumps and translationally unstable static bumps.
To predict different phases in the parameter space, however,
we need to study the stability of fixed point solutions given by
Eqs. (\ref{eq:dukkdt}), (\ref{eq:dpkkdt}) and (\ref{eq:dBdt}).
A detailed discussion of the stability issue can be found in Appendix
\ref{subsec:moving_std}.

By studying the stability of fixed point moving solutions, we can 
obtain a phase boundary separating silent and 
moving phases, as plotted as a dashed line in Figure 
\ref{fig:phase_diagram_std}.  In Figure \ref{fig:phase_diagram_std},
there is a rather complete phase diagram for the model of 2D CANNs 
with STD we are studying here.  For small enough $\widetilde\beta$s 
(below the solid curve in Figure \ref{fig:phase_diagram_std}), 
since static bumps are stable in amplitude and translation, 
this parameter region supports only static bumps. This region
is called the static phase.

For parameters between the dotted line and the solid line,
static bumps are stable in amplitude but unstable against translational
distortions. So in this case, in a rather short time window,
both static bumps and moving bumps are able to exist. Because
once a translationally unstable static bump get perturbed, it becomes 
a moving bump.  Since both static bumps and moving bumps are 
observed in this parameter region, we call it the 
bistable phase. If we further increase $\widetilde\beta$, however,
even static bump cannot be observed.  And only 
moving bumps can be observed in the parameter region, we call it 
the moving phase. When both $\widetilde k$ and $\widetilde \beta$ are 
large, no non-trivial solution can be found in our analysis or 
simulations. 

In this study we did not find the homogeneous firing patterns
reported by \cite{York2009}. In the 1D case, however, homogeneous 
firing can be found in the $\widetilde k \rightarrow 0$ regime
\citep{Wang2014}. Since we are focusing on moderate magnitudes of
$\widetilde k$, the possibility of uniform firing behavior in 
2D CANNs with STD near $\widetilde k = 0$ is reserved for future
studies.

\section{CANNs with SFA}

\subsection{Stationary Solution}
\begin{figure}
\begin{centering}
\includegraphics[width=0.8\textwidth]{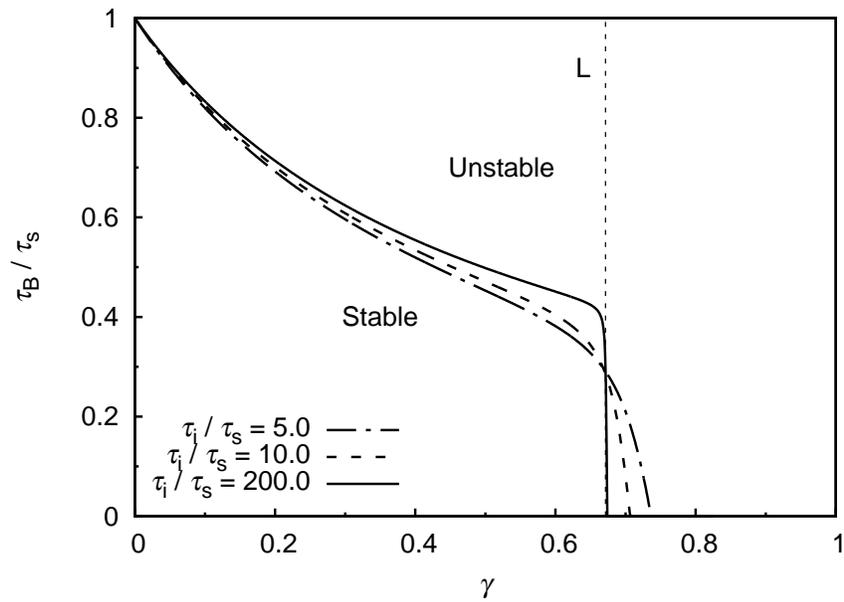}\protect\caption{\label{fig:max_tauB} Maximum $\tau_{B}/\tau_{s}$ able to stabilize
the amplitude of stationary solutions as a function of $\gamma$,
given that $\widetilde{k}=0.3$. $L$: maximum $\gamma$ of the existence
of stationary solutions. }

\par\end{centering}

\end{figure}

By setting $\widetilde\beta=0$ in Eq. (\ref{eq:dpdt}), the
CANN model we are considering becomes a network with 
SFA only.  To study the stationary solution in this case, we assume
\begin{align}
u\left(\mathbf{x}\right)  = & ~ u_{00}\exp\left(-\frac{\mathbf{x}^{2}}{4a^{2}}\right),\label{eq:u0_sub_stat_fixed}\\
v\left(\mathbf{x}\right)  = & ~ v_{00}\exp\left(-\frac{\mathbf{x}^{2}}{4a^{2}}\right).\label{eq:v0_sub_stat_fixed}
\end{align}
By substituting Eqs. (\ref{eq:u0_sub_stat_fixed}) and 
(\ref{eq:v0_sub_stat_fixed}) into Eqs. (\ref{eq:dudt}) and 
(\ref{eq:dvdt_sub}), we found that the fixed point solution 
is given by
\begin{align}
\widetilde{u}_{00}  = & ~ 4\frac{1\pm\sqrt{1-\left(1+\gamma\right)^{2}\widetilde{k}}}{\left(1+\gamma\right)\widetilde{k}},\label{eq:u0_sub_stat}\\
\widetilde{v}_{00}  = & ~ 4\gamma\frac{1\pm\sqrt{1-\left(1+\gamma\right)^{2}\widetilde{k}}}{\left(1+\gamma\right)\widetilde{k}},\label{eq:v0_sub_stat}\\
B  = & ~ 2\frac{1\pm\sqrt{1-\left(1+\gamma\right)^{2}\widetilde{k}}}{\left(1+\gamma\right)^{2}\widetilde{k}}.\label{eq:B_sub_stat}
\end{align}
$\widetilde{u}_{00}$ is a rescaled variable defined by $\widetilde{u}_{00}\equiv\rho J_{0}u_{00}$.
Similarly, $\widetilde{v}_{00}\equiv\rho J_{0}v_{00}$. We can also rescale
$u(\mathbf{x},t)$ and $v(\mathbf{x},t)$ in the same way: i.e., $\widetilde{u}(\mathbf{x},t)\equiv\rho J_{0}u(\mathbf{x},t)$
and $\widetilde{v}(\mathbf{x},t)\equiv\rho J_{0}v(\mathbf{x},t)$. For
$\tau_{B}=0$, the fixed point solution with the larger $\widetilde{u}_{00}$
and $\widetilde{v}_{00}$ is stable whenever 
\begin{equation}
0<\widetilde{k}<\frac{1+2\gamma}{\left(1+\gamma\right)^4},\label{eq:amp_stable_sub}
\end{equation}
which is labeled by curve $L$ in Figure \ref{fig:max_tauB} for 
$\widetilde{k}=0.3$. The stability issue can be studied by 
considering the dynamics of distortions of dynamical variables 
from their fixed point solutions. Let 
\begin{align}
\widetilde{u}\left(\mathbf{x},t\right) = & ~ \left[\widetilde{u}_{00}+\delta\widetilde{u}_{00}\left(t\right)\right]\exp\left(-\frac{\mathbf{x}^{2}}{4a^{2}}\right),\\
\widetilde{v}\left(\mathbf{x},t\right) = & ~ \left[\widetilde{v}_{00}+\delta\widetilde{v}_{00}\left(t\right)\right]\exp\left(-\frac{\mathbf{x}^{2}}{4a^{2}}\right),\\
B\left(t\right) = & ~ B+\delta B\left(t\right).
\end{align}
By studying the stability of fixed point solutions,
we obtain a parameter region for static bumps with stable amplitudes.
Detailed analysis of the stability issue can be found in Appendix
\ref{subsec:static_sub}. Regions for $\widetilde k = 0.3$ and 
various $(\tau_i/\tau_s)$s are 
shown in Figure \ref{fig:max_tauB}.
For $\tau_{B}>0$, the parameter region over the $(\widetilde{k},\gamma)$
space becomes smaller as $\tau_{B}$ increases. At the 
$\tau_{i}\gg\tau_{s}$ limit, the range of $\widetilde{k}$ to stabilize 
the static solution is given by 
\begin{equation}
0<\widetilde{k}<\frac{2\gamma+1}{\left(1+\gamma\right)^{4}},
\end{equation}
which agrees with Eq. (\ref{eq:amp_stable_sub}).
On the other hand, if $\tau_{B}/\tau_{s}\ge1$, there will be no static
solution. In Figure \ref{fig:max_tauB}, we have shown that if $\gamma$
becomes larger, the maximum value of $\tau_{B}/\tau_{s}$ to stabilize
the static solution will be smaller.

\subsection{Translational Stability}

To study the translational stability, we consider the lowest-order 
asymmetric distortions added to the stationary solution. The major 
concern here is the step function in Eq. (\ref{eq:f_u2v}), but the 
step function does not have a first-order effect on translational 
distortions. We let
\begin{align}
u\left(\mathbf{x},t\right) = & ~ u_{00}\left(t\right)\exp\left(-\frac{\left|\mathbf{x}\right|^{2}}{4a^{2}}\right),\label{eq:u0_3}\\
v\left(\mathbf{x},t\right) = & ~ v_{00}\left(t\right)\exp\left(-\frac{\left|\mathbf{x}\right|^{2}}{4a^{2}}\right)+v_{10}\left(t\right)\frac{x_{0}}{a}\exp\left(-\frac{\left|\mathbf{x}\right|^{2}}{4a^{2}}\right).\label{eq:v0v1}
\end{align}
As we are studying the stability issue of the static solution,
let us adopt the solution in Eqs. (\ref{eq:u0_sub_stat}) - (\ref{eq:B_sub_stat})
and put $v_{10}=0$. Then the dynamics of distortions of dynamical 
variables near fixed point static solutions becomes
\begin{equation}
 \tau_{s}\frac{d}{dt}\left(\begin{array}{c}
\delta\widetilde{u}_{00}\left(t\right)\\
\delta\widetilde{v}_{00}\left(t\right)\\
\delta B\left(t\right)\\
\delta\widetilde{v}_{10}\left(t\right)
\end{array}\right)\\
 = \left(\begin{array}{cc}
\mathscr{A}_{\rm SFA} & 0\\
 0 & -\frac{\tau_{s}}{\tau_{i}}+\gamma
\end{array}\right)
 \left(\begin{array}{c}
\delta\widetilde{u}_{00}\left(t\right)\\
\delta\widetilde{v}_{00}\left(t\right)\\
\delta B\left(t\right)\\
\delta\widetilde{v}_{10}\left(t\right)
\end{array}\right) ,
\label{eq:trans_stable_sub}
\end{equation}
where the $3\times3$ matrix $\mathscr{A}_{\rm SFA}$ 
is provided in Appendix \ref{subsec:static_sub}.
$\mathscr{A}_{\rm SFA}$ determines amplitudal stability of 
the static solution. Clearly, the variable $v_{10}$ becomes unstable 
if $\gamma>\tau_{s}/\tau_{i}$. This implies that if the dynamics of SFA is
slow enough, the static solution will be translationally unstable.

\subsection{Moving Solution}

Once $\gamma>\tau_{s}/\tau_{i}$ is satisfied, spontaneous motion
of a bump-shaped solution in 2D CANN with SFA becomes possible. As 
in the STD case, however, lower-order expansions of $u\left(\mathbf{x},t\right)$
and $p\left(\mathbf{x},t\right)$ are not sufficient to describe the
moving solutions. We have to consider higher-order expansions if we want
to obtain good predictions on the behavior of moving solutions. 
Surprisingly, we found that a limited order of expansion of 
$u\left(\mathbf{x},t\right)$ and $v\left(\mathbf{x},t\right)$ 
can give some fairly good predictions on the moving solutions.

In general, we consider
\begin{align}
u\left(\mathbf{x},t\right) = & ~ \sum_{k_{0}k_{1}}u_{k_{0}k_{1}}\left(t\right)\phi_{k_{0}}\left(\xi_{0}\right)\phi_{k_{1}}\left(\xi_{1}\right),\\
v\left(\mathbf{x},t\right) = & ~ \sum_{k_{0}k_{1}}v_{k_{0}k_{1}}\left(t\right)\phi_{k_{0}}\left(\xi_{0}\right)\phi_{k_{1}}\left(\xi_{1}\right).
\end{align}
Here 
\begin{equation}
\phi_{k_{i}}\left(\xi_{i}\right)=\frac{1}{\sqrt{\sqrt{2\pi}ak_{i}!}}H_{k_{i}}\left(\frac{\xi_{i}}{a}\right)\exp\left(-\frac{\xi_{i}^{2}}{4a^{2}}\right)
\end{equation}
and $H_{k_{i}}$ is the $k_{i}^{\text{th}}$-order probabilist's Hermite
polynomial. Eqs. (\ref{eq:u0_sub_stat}) and (\ref{eq:v0_sub_stat})
become
\begin{align}
   &~ \tau_{s}\frac{d\widetilde{u}_{k_{0}k_{1}}}{dt}\left(t\right) \nonumber \\
 = &~ -\widetilde{u}_{k_{0}k_{1}}\left(t\right)-\widetilde{v}_{k_{0}k_{1}}\left(t\right)+\frac{1}{B\left(t\right)}\sum_{n_{0}n_{1}m_{0}m_{1}}C_{n_{0}m_{0}}^{k_{0}}C_{n_{1}m_{1}}^{k_{1}}\widetilde{u}_{n_{0}n_{1}}\left(t\right)\widetilde{u}_{m{}_{0}m_{1}}\left(t\right), \nonumber\\
  & ~ \qquad\qquad\qquad -\frac{\tau_{s}c_{0}}{2a}\left[\sqrt{k_{0}}\widetilde{u}_{k_{0}-1,k_{1}}\left(t\right)-\sqrt{k_{0}+1}\widetilde{u}_{k_{0}+1,k_{1}}\left(t\right)\right]\nonumber \\
   &~ \qquad\qquad\qquad-\frac{\tau_{s}c_{1}}{2a}\left[\sqrt{k_{1}}\widetilde{u}_{k_{0},k_{1}-1}\left(t\right)-\sqrt{k_{1}+1}\widetilde{u}_{k_{0},k_{1}+1}\left(t\right)\right]\\
   &~ \tau_{i}\frac{d\widetilde{v}_{k_{0}k_{1}}}{dt}\left(t\right)  \nonumber \\
= &~ -\widetilde{v}_{k_{0}k_{1}}\left(t\right)+\gamma\widetilde{u}_{k_{0}k_{1}}\left(t\right) \nonumber \\
&~\qquad\qquad\qquad-\frac{\tau_{i}c_0}{2a}\left[\sqrt{k_{0}}\widetilde{v}_{k_{0}-1,k_{1}}\left(t\right)-\sqrt{k_{0}+1}\widetilde{v}_{k_{0}+1,k_{1}}\left(t\right)\right]\nonumber \\
&~\qquad\qquad\qquad-\frac{\tau_{i}c_{1}}{2a}\left[\sqrt{k_{1}}\widetilde{v}_{k_{0},k_{1}-1}\left(t\right)-\sqrt{k_{1}+1}\widetilde{v}_{k_{0},k_{1}+1}\left(t\right)\right]
\end{align}

Together with the self-consistent condition, Eq. (\ref{eq:self-con}),
the fixed point solution is solvable. Here the definition of $C_{nm}^k$ 
is not the same as that in Eq. (\ref{eq:std_Cnmk}), which is defined by
\begin{equation}
C_{nm}^{k}  \equiv  \int d\xi\phi_{k}\left(\xi\right)\int d\xi^{\prime}\frac{1}{\sqrt{2\pi}a}e^{-\frac{\left(\xi-\xi^{\prime}\right)^{2}}{2a^{2}}}\phi_{n}\left(\xi^{\prime}\right)\phi_{m}\left(\xi^{\prime}\right). \label{eq:Cknm_sub}
\end{equation}
$C_{00}^0$ can be calculated explicitly. In general, $C^k_{nm}$ can be 
obtained by using the recurrence relations given in Appendix
\ref{subsec:moving_Cknm_sub}.

\begin{figure}
\begin{centering}
\includegraphics[width=0.8\textwidth]{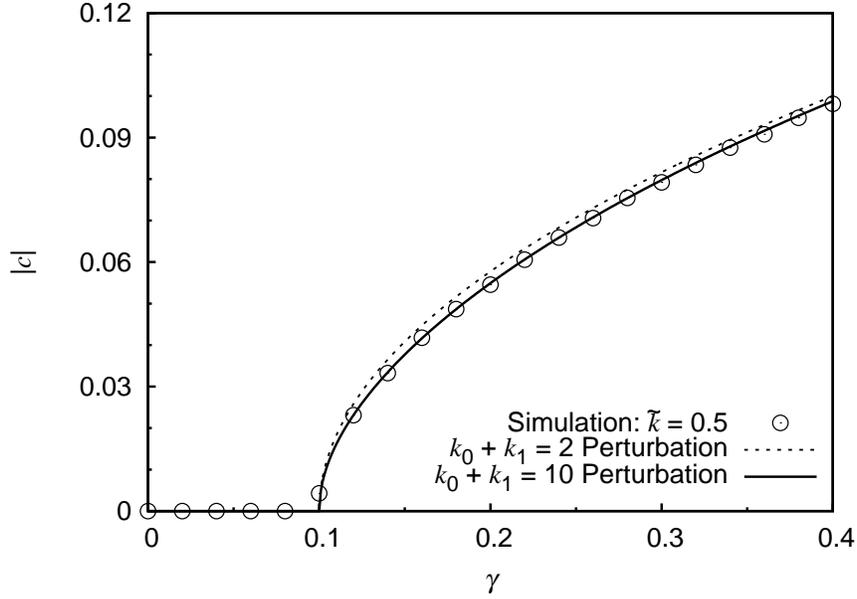}
\par\end{centering}

\centering{}\protect\caption{\label{fig:sub_c} Speed of spontaneous motions as a function of $\gamma$.}
\end{figure}

If we consider only terms up
to $k_{0}+k_{1}=2$ and motion along the $x_{0}$-direction, we can obtain
the intrinsic speed of the moving solution (detailed derivation can be 
found in Appendix \ref{subsec:moving_speed_sub}):
\begin{equation}
\frac{\tau_{s}\left|c_{0}\right|}{2a}=\frac{\tau_{s}}{\tau_{i}}\sqrt{\frac{1}{3}\left(\frac{\tau_{i}}{\tau_{s}}\gamma-1\right)}.
\end{equation}
As the preferred stimulus space is rotationally symmetric,
this speed is applicable to motion in any direction. Even though this
is an approximated solution with relatively few terms, the prediction
on the intrinsic speed is fairly good, as shown in Figure \ref{fig:sub_c}.

\begin{figure}
\begin{centering}
\includegraphics[width=\textwidth]{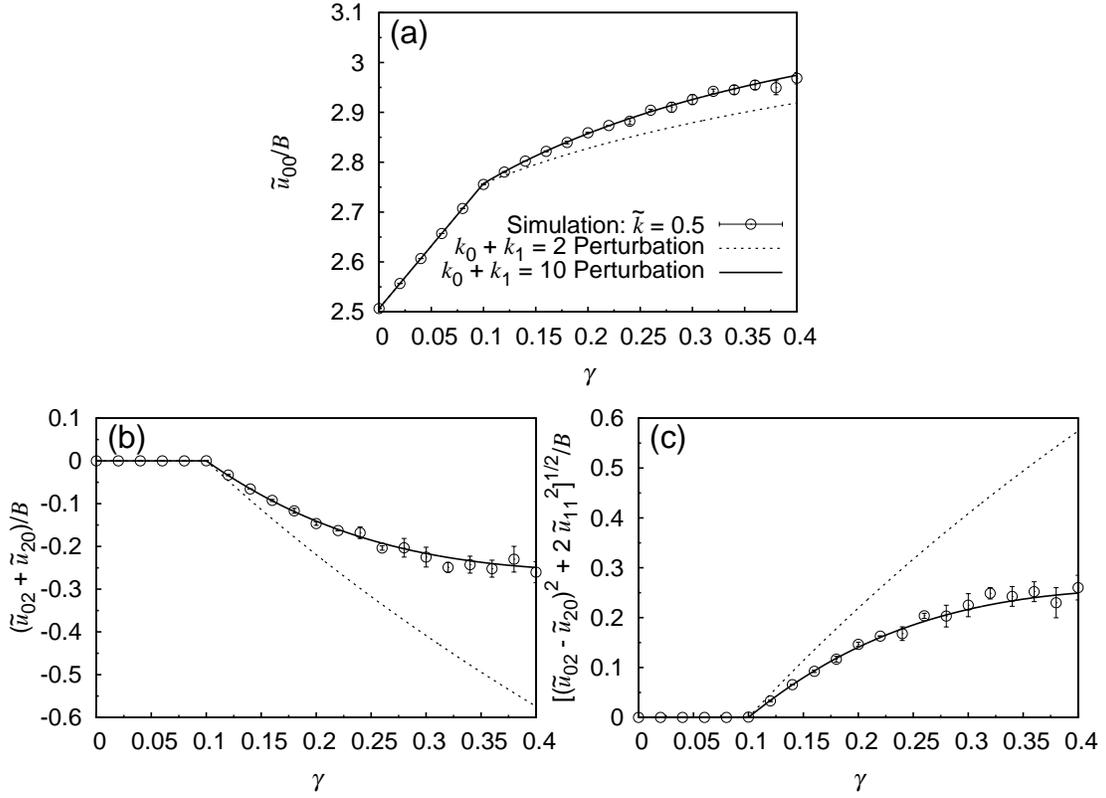}
\par\end{centering}

\protect\caption{\label{fig:sub_u00} 
(a) Projection of $\widetilde{u}\left(\mathbf{x},t\right)$
on $\phi_{0}\left(\xi_{0}\right)\phi_{0}\left(\xi_{1}\right)$. 
(b) Projection of $\widetilde{u}\left(\mathbf{x},t\right)$
on rotationally symmetric basis function with $k_{0}+k_{1}=2$.
(c) Projection of $\widetilde{u}\left(\mathbf{x},t\right)$
on anisotropic basis function with $k_{0}+k_{1}=2$.
(a) - (c) Parameters: $a=0.5$, $\tau_{i}/\tau_{s}=10$ and $\tau_{B}/\tau_{s}=0.1$.}
\end{figure}

This result suggests that, in this case, the terms with $k_{0}+k_{1}\le2$
are sufficient to give some good predictions on the behavior of bump-shaped
solutions. For predictions on measurements of different components
of the moving bump, however, we still need to use higher-order perturbation. 
In Figures \ref{fig:sub_u00}(a)~-~(c), we have shown that higher-order 
perturbation is needed to predict $\widetilde{u}_{k_{0}k_{1}}/B$
if the bump is moving. As in Figures \ref{fig:std_u00}(b) 
and \ref{fig:std_u00}(c),
in Figures \ref{fig:sub_u00}(b) and \ref{fig:sub_u00}(c), we do not compare
$\widetilde{u}_{20}$, $\widetilde{u}_{02}$ and $\widetilde{u}_{11}$ directly.
Since a moving bump can move in any directions, we
compare their projections in polar coordinates. Let us consider 
$\xi_{0}=r\cos\theta$ and $\xi_{1}=r\sin\theta$.  Then we have
\begin{align}
   & ~ \widetilde{u}_{20}\phi_{2}\left(\xi_{0}\right)\phi_{0}\left(\xi_{1}\right)+\widetilde{u}_{02}\phi_{0}\left(\xi_{0}\right)\phi_{2}\left(\xi_{1}\right)+\widetilde{u}_{11}\phi_{1}\left(\xi_{0}\right)\phi_{1}\left(\xi_{1}\right)\nonumber \\
  = & ~ \left(\widetilde{u}_{20}+\widetilde{u}_{02}\right)\frac{1}{2a\sqrt{\pi}}\left(\frac{1}{2}\frac{r^{2}}{a^{2}}-1\right)e^{-\frac{r^{2}}{4a^{2}}}+\left(\widetilde{u}_{20}-\widetilde{u}_{02}-\sqrt{2}i\widetilde{u}_{11}\right)\frac{1}{8a\sqrt{\pi}}\frac{r^{2}}{a^{2}}e^{-\frac{r^{2}}{4a^{2}}}e^{i2\theta}\nonumber \\
   & ~ +\left(\widetilde{u}_{20}-\widetilde{u}_{02}+\sqrt{2}i\widetilde{u}_{11}\right)\frac{1}{8a\sqrt{\pi}}\frac{r^{2}}{a^{2}}e^{-\frac{r^{2}}{4a^{2}}}e^{-i2\theta}.\label{eq:sfa_rect2rot}
\end{align}
In Figures \ref{fig:sub_u00}(b) and \ref{fig:sub_u00}(c), we compare predictions
and simulation results of $\left(\widetilde{u}_{20}+\widetilde{u}_{02}\right)$
and $\left|\widetilde{u}_{20}-\widetilde{u}_{02}-\sqrt{2}i\widetilde{u}_{11}\right|$
for $k_{0}+k_{1}=2$ and $k_0+k_1=10$. $\left(\widetilde{u}_{20}+\widetilde{u}_{02}\right)$
is the average change in the width of the bump-shaped profile, while $\left|\widetilde{u}_{20}-\widetilde{u}_{02}-\sqrt{2}i\widetilde{u}_{11}\right|$
is the magnitude of the anisotropic mode. The graphical illustration
of Eq. (\ref{eq:sfa_rect2rot}) is omitted because it is similar to the 
illustrations shown in Figure \ref{fig:std_basis_polar}.

The behavior of moving solutions of CANNs with SFA
is similar to that of moving solutions of 
CANNs with STD. For $\widetilde{u}_{00}/B$, its
trend is basically the same as $\widetilde{u}_{00}/B$ in the case with
STD shown in Figure \ref{fig:std_u00}. The transition happens whenever
$\gamma=\tau_{s}/\tau_{i}$. Also, anisotropic modes will be
available only when the bump is moving, while the width of the $u(\mathbf{x},t)$
profile increases as the strength of SFA
increases. The behavior of the average change in width is similar
to the behavior of the change in height, $\widetilde{u}_{00}$. The dependence
of anisotropic modes on the strength of SFA, $\gamma$,
implies that the widening effect is not uniform when the bump is moving,
which is similar to what is seen in a 2D CANN with STD.

\begin{figure}
\begin{centering}
\includegraphics[width=0.8\textwidth]{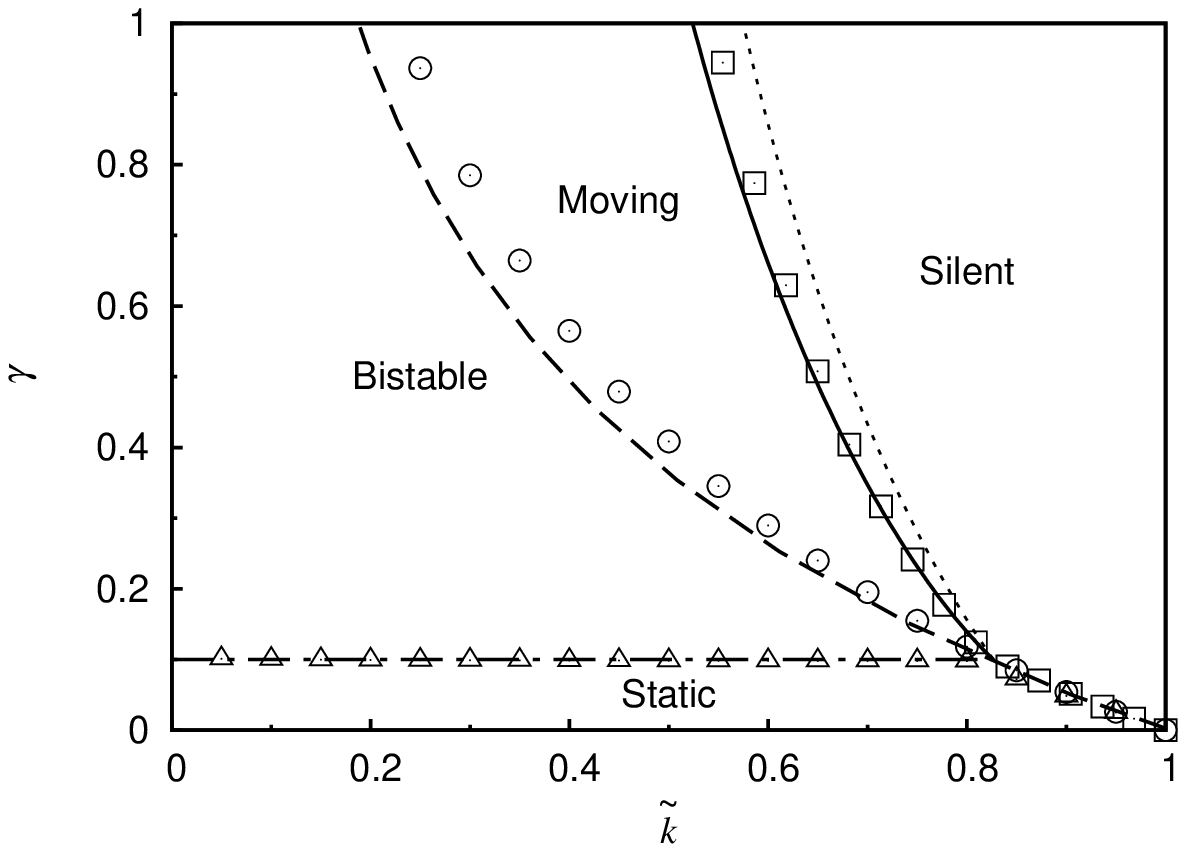}
\par\end{centering}

\begin{centering}
\protect\caption{\label{fig:sub_phase_diagram} Phase diagram over the parameter space
spanned by $\left(\widetilde{k},\gamma\right)$. Dotted line: boundary predicted
for moving solutions by $k_{0}+k_{1}=2$ perturbation. Solid line: boundary predicted 
for moving solutions by $k_{0}+k_{1}=10$ perturbation. Dotted line: boundary predicted
for static solutions with stable amplitudes by $k_{0}+k_{1}=0$ perturbation. Dot-dashed
line: boundary predicted for static solutions with stable amplitudes
and translational stability by $k_{0}+k_{1}=0$ perturbation. 
Parameters: same as Figure \ref{fig:sub_u00}. Symbols:
corresponding simulations.}
\end{centering}
\end{figure}

\subsection{Phase Diagram}

A phase diagram similar to Figure \ref{fig:phase_diagram_std} for SFA 
can be obtained in a similar manner used for CANNs with STD.  In the 
previous subsection, we have shown that perturbative analysis is 
able to predict dynamical variables of the model. By solving for the moving 
solutions numerically and testing their stability, we can predict
the phase boundary separating parameter regions for moving solutions 
and trivial solution (silent phase).  Using it together with the 
stability conditions for static bumps given in Eqs. 
(\ref{eq:amp_stable_sub}) and (\ref{eq:trans_stable_sub}),
we can predict the phase diagram for SFA. We found that the predicted 
phase diagram matches the phase diagram obtained from computer 
simulations (Figure \ref{fig:sub_phase_diagram}).  
As in Figure \ref{fig:phase_diagram_std}, there are four phases in 
the phase diagram: a static phase, a bistable phase, a
moving phase and a silent phase.  Their meanings are the same as those of their
counterparts in Figure \ref{fig:phase_diagram_std}.
Remarkably, expansions up to $k_{0}+k_{1}=2$ can also predict the 
phase diagram well, which is similar to the prediction of 
intrinsic speed of moving bumps in CANNs with SFA.

\section{Discussion}

\subsection{Intrinsic Phases and Phase Diagrams}

In the parameter spaces of the two models we studied in this paper, 
there are static, bistable, moving and silent phases. When 
$\widetilde\beta$ is large enough, moving bumps can 
be found in bistable phase and moving phase. This behavior can
also be seen in the case of SFA. In both STD and SFA cases, 
whenever $\widetilde\beta$ or $\gamma$ is not too large (within
the bistable phase), static bumps can still exist for a not insignificant
period of time.  If $\widetilde\beta$ or $\gamma$ is too large,
only trivial solutions are stable.

Behaviors of CANNs with STD are very similar to those of CANNs with SFA.  
This suggests that multiplicative dynamical suppression mechanisms 
(represented by STD) and subtractive dynamical suppression mechanisms 
(represented by SFA) can generate similar intrinsic dynamics, 
especially with regard to spontaneous motion.
It also implies that other smooth models having homogeneous couplings 
and dynamical suppression mechanisms should have a similar phase diagram 
consisting of static, moving, bistable and silent phases.

In some studies on CANNs with STD or SFA it is reported that, 
with some parameters, uniform firing pattern can be found in 
the network \citep[e.g.][]{York2009}.
In the present study, uniform firing is not included because the 
interactive range of the global inhibition in this model is infinite.
So uniform firing may only be found in a tight parameter region near 
$\widetilde k\rightarrow 0$. For the 1D STD case, uniform firing can be 
found in a region $\widetilde\beta \rightarrow 0$ and 
$\widetilde k \rightarrow 0$ \citep{Wang2014}.  Uniform firing
should also be found in 2D CANNs with STD or SFA, as should
spiral waves and breathing wavefronts. There is a report on 
spiral waves and breathing wavefronts in 2D 
CANNs with STD \citep[e.g.][]{Kilpatrick2010}, but those phenomena are
missing from the phase diagram we predicted because spiral 
waves and breathing wavefronts are not local
patterns. Therefore in the present model they may exist if 
the magnitude of divisive global inhibition is very small.
Richer dynamics of the present model for $\widetilde k\rightarrow 0$
is reserved for future investigations.

\subsection{Effectiveness of Perturbative Approach}

In this paper we have shown that, in the present particular model,
the perturbative expansion method is applicable to the study of
continuous attractor neural networks (CANNs) with short-term synaptic
depression (STD) or spike frequency adaptation (SFA). We found in 
this study that both STD and SFA can drive similar
intrinsic dynamics of the local neural activity profile. In Figures
\ref{fig:phase_diagram_std} and \ref{fig:sub_phase_diagram} there
are four phases: silent, static, bistable and moving. Using perturbative
expansions on the dynamical variables, we can successfully predict
the phase diagrams in both cases with STD and SFA.

As expected, with low-order expansions (i.e., when $k_{0}+k_{1}$ is 
small), predictions on $\widetilde{u}_{k_{0}k_{1}}$'s of the static
solutions work well in cases with STD and SFA.
For moving solutions, higher-order expansions are needed to obtain
more accurate solutions. In Figures \ref{fig:std_u00}(a) - (c)
and Figure \ref{fig:sub_u00} it is shown that low-order perturbative 
expansions, $k_{0}+k_{1}=3$ for STD and $k_{0}+k_{1}=2$
for SFA, are able to show the general trend of dynamical variables. 
Especially, the second-order transitions in all the dynamical variables
(i.e. discontinuities of slopes)
are observed at this level of perturbative expansion. This suggests 
that low-order perturbative expansions are good enough for studying 
general phenomena in different phases.

To predict the behavior of the dynamical variables more accurately,
we need to use higher-order perturbative expansions. We have shown
in Figures \ref{fig:std_u00}(a) - (c) and Figure \ref{fig:sub_u00}
that higher-order perturbative expansions, $k_{0}+k_{1}=11$
for STD and $k_{0}+k_{1}=10$ for SFA, can fit measurements from
simulations accurately. Higher-order terms do not, however, significantly
improve prediction of the speed of spontaneous motion. This suggests
that lower-order perturbative modes are most important to the motion
of the local neural activity profile.

\begin{figure}
\begin{centering}
\includegraphics[width=1\textwidth]{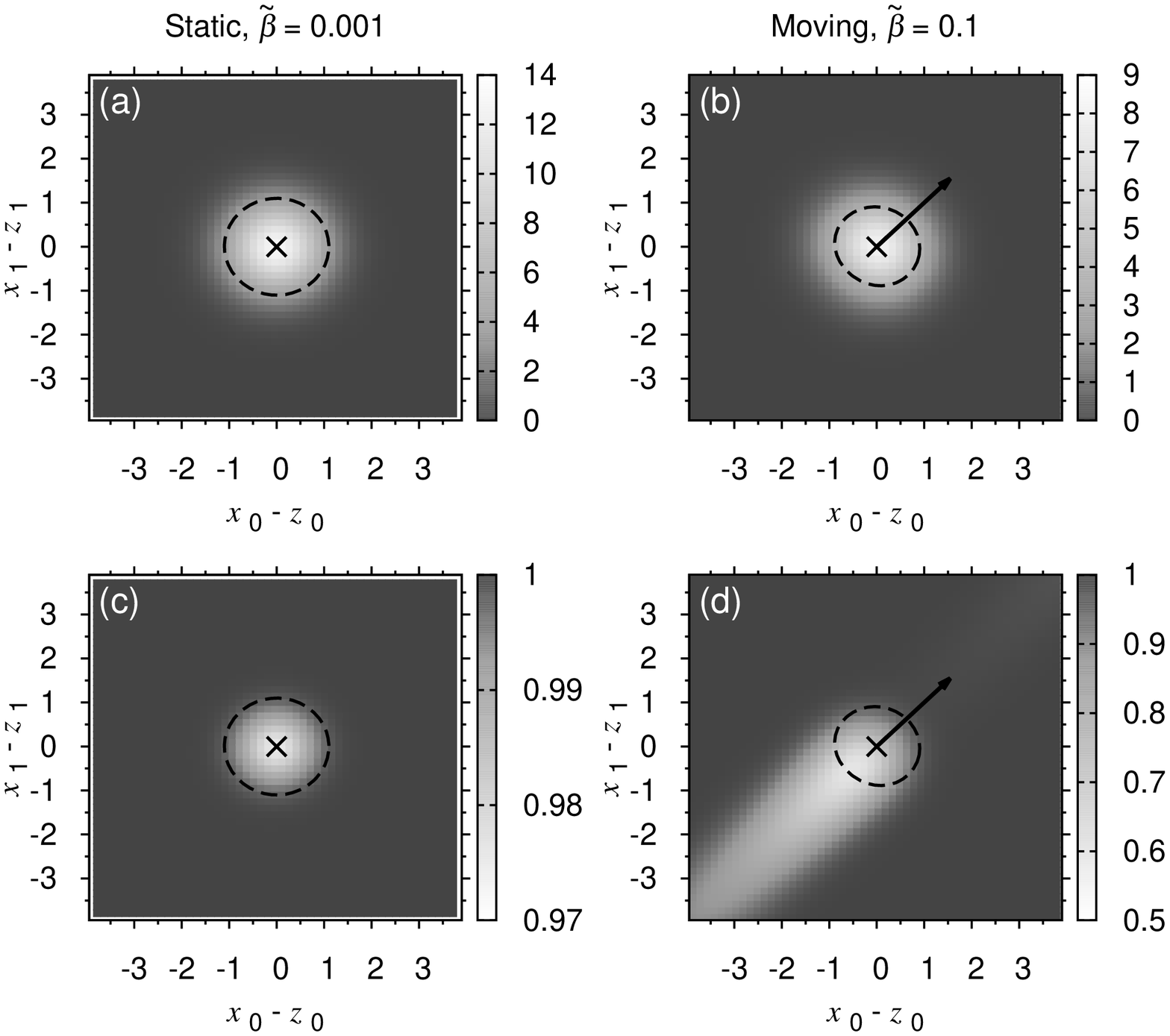}
\par\end{centering}

\protect\caption{\label{fig:std_exp} Snapshots of $\widetilde{u}\left(\mathbf{x},t\right)$
and $p\left(\mathbf{x},t\right)$ in the static phase ((a) and (c)) and
moving phase ((b) and (d)). (a) and (b) are $\widetilde{u}\left(\mathbf{x},t\right)$,
while (c) and (d) are $p\left(\mathbf{x},t\right)$. Dashed lines:
contours for $\widetilde{u}\left(\mathbf{x},t\right)=4$. $\left(z_{0},z_{1}\right)$
is the center of mass of $\widetilde{u}\left(\mathbf{x},t\right)$. The moving
direction of the bump is the $\left(1,1\right)$ direction, shows by the arrows
in (b) and (d). Parameters: $\widetilde{k}=0.5$, $\tau_{d}/\tau_{s}=50$ and
$\widetilde{\beta}=0.001$ for (a) and (c), and $\widetilde{\beta}=0.1$ for (b) and (d).}
\end{figure}

\subsection{Asymmetric Modes and Moving Solutions}

Higher-order perturbative modes are essential for predicting the behavior
of dynamical variables of moving solutions accurately because the dynamical
variables will become asymmetric about the center of mass of $\widetilde{u}\left(\mathbf{x},t\right)$
if the neural activity profile is moving. In Figure \ref{fig:std_exp},
there are two examples of CANNs with different levels of STD. Figures
\ref{fig:std_exp}(a) and \ref{fig:std_exp}(c) are with $\widetilde{\beta}=0.001$,
which corresponds to the static phase. In this case, both the average membrane potential profile $\widetilde{u}\left(\mathbf{x},t\right)$ and the 
fraction of available neurotransmitters $p\left(\mathbf{x},t\right)$ are rotationally
symmetric about the center of mass of $\widetilde{u}\left(\mathbf{x},t\right)$.
Therefore, in the static phase, low-order perturbation is good enough for making predictions.

Snapshots of $\widetilde{u}\left(\mathbf{x},t\right)$ and
$p\left(\mathbf{x},t\right)$ in the moving phase
are shown in Figures \ref{fig:std_exp}(b) and \ref{fig:std_exp}(d).
Here the level of STD is $\widetilde{\beta}=0.1$. In this case, $\widetilde{u}\left(\mathbf{x},t\right)$
seems to be rotationally symmetric about its center of mass. Actually,
the shape of the Gaussian-like average membrane potential profile is slightly
squeezed along the moving direction. For $p\left(\mathbf{x},t\right)$, however,
the deformation of the shape is more significant. In Figure \ref{fig:std_exp}(d),
the profile of $p\left(\mathbf{x},t\right)$ is strongly biased against
the moving direction. This highly skewed profile of $p\left(\mathbf{x},t\right)$
makes high-order perturbative mode more important, especially for
cases involving stability and predictions of dynamical variables.

\subsection{Limitation of Perturbative Approach}

In this paper, we have shown the perturbative approach is able to
successfully predict 1) the phase diagrams of CANNs with STD or SFA,
2) the speed of spontaneous motion and 3)
the dynamical variables (e.g. $\widetilde{u}\left(x,t\right)$). 
Phenomena richer than spontaneous motion of local neural activity,
for example, breathing wavefronts on neural networks have been 
considered theoretically \citep{Kilpatrick2010}. The perturbative 
formulation is not applicable in that case, however, because 
the basis function we used here are local, while the traveling 
wavefronts are globally spreading. For the same reason, this method cannot
be used to analyze also spiral waves on a 2D field.

In the case of collisions of two moving bump-shaped profiles of neuronal
activity, one may find the dynamics hard to analyze by the 
perturbative method. This is because there are two centers of mass, 
one for each bump, and this makes the choice of the origin of basis functions
confusing. For example, if one chose one of the centers of mass to
be the origin of basis functions, the chosen family of basis functions
will not be optimal for the other bump and will make the expansion of
that bump less efficient.

\section*{Conclusion}

In this paper we studied models with short-term synaptic depression
and spike frequency adaptation and these models which was based on 
a two-dimensional CANN model with divisive global inhibition. 
We found that their intrinsic dynamics were similar. First, there 
are four phases in each scenario. They are static, moving, bistable 
and silent. The moving phase is the phase in which the bump-shaped 
profile moves spontaneously, while the static phase is the phase in
which the bump-shaped profile cannot move spontaneously. Interestingly,
in the bistable phase, the CANN can support both moving profiles and
static profiles.

Second, there are clear phase transitions between static solutions
and moving solutions. In Figures \ref{fig:std_u00}, \ref{fig:sub_c} and
\ref{fig:sub_u00} there is a clear discontinuity in slope between the
two states. This suggests that, spontaneous motion can affect the
shapes of the bump-shaped profiles.

The stability of steady states, the shapes of the dynamical variables
and the speed of spontaneous motion can be predicted by perturbative
analysis, but perturbative method cannot be used in some
scenarios including spiral waves, breathing wavefronts and collisions
of multiple bumps in 2D fields.

\subsection*{Acknowledgments}

This study is partially supported by the Research Grants Council of
Hong Kong (grant numbers 604512, 605813 and N\_HKUST 606/12).

\renewcommand\thesubsection{\Alph{subsection}}
\numberwithin{equation}{subsection}
\setcounter{subsection}{0}
\section*{Appendix}

\subsection{\label{subsec:static_std} Amplitudal Stability of Bumps on 2D CANNs with STD}

To study the stability of fixed point solutions of 
Eqs. (\ref{eq:du00dt_std}) - (\ref{eq:dBdt_std}), we assume
\begin{align}
\widetilde u_{00}\left(t\right) = & ~ \widetilde u_{00}^{*} + \delta \widetilde u_{00}\left(t\right),\\
p_{00}\left(t\right)  = & ~ p_{00}^{*} + \delta p_{00}\left(t\right),\\
B\left(t\right)  = & ~ B^{*} + \delta B\left(t\right).
\end{align}
Linearizing Eqs. (\ref{eq:du00dt_std}) - (\ref{eq:dBdt_std}), 
we obtain
\begin{equation}
\tau_{s}
\frac{d}{dt}
\left(\begin{array}{c}
\widetilde{u}_{00}\left(t\right) \\
p_{00}\left(t\right) \\
B \left(t\right) 
\end{array}\right)
=
\mathscr{A}_{\rm STD}
\left(\begin{array}{c}
\widetilde{u}_{00}\left(t\right) \\
p_{00}\left(t\right) \\
B \left(t\right) ,
\end{array}\right)
\end{equation}
where 
\begin{align}
\mathscr{A}_{\rm STD}  = & ~ \left(\begin{array}{ccc}
\frac{\partial F_{\widetilde{u}_{00}}\left(\widetilde{u}_{00},p_{00},B\right)}{\partial\widetilde{u}_{00}} & \frac{\partial F_{\widetilde{u}_{00}}\left(\widetilde{u}_{00},p_{00},B\right)}{\partial p_{00}} & \frac{\partial F_{\widetilde{u}_{00}}\left(\widetilde{u}_{00},p_{00},B\right)}{\partial B}\\
\frac{\partial F_{p_{00}}\left(\widetilde{u}_{00},p_{00},B\right)}{\partial\widetilde{u}_{00}} & \frac{\partial F_{p_{00}}\left(\widetilde{u}_{00},p_{00},B\right)}{\partial p_{00}} & \frac{\partial F_{p_{00}}\left(\widetilde{u}_{00},p_{00},B\right)}{\partial B}\\
\frac{\partial F_{B}\left(\widetilde{u}_{00},p_{00},B\right)}{\partial\widetilde{u}_{00}} & \frac{\partial F_{B}\left(\widetilde{u}_{00},p_{00},B\right)}{\partial p_{00}} & \frac{\partial F_{B}\left(\widetilde{u}_{00},p_{00},B\right)}{\partial B}
\end{array}\right)\label{eq:A_std}\\
  = & ~ \left(\begin{array}{ccc}
-1+\frac{1}{B}\widetilde{u}_{00}\left(1-\frac{4}{7}\widetilde{p}_{00}\right) & -\frac{2}{7}\frac{1}{B}\widetilde{u}_{00}^{2} & -\frac{1}{2}\frac{1}{B^{2}}\widetilde{u}_{00}^{2}\left(1-\frac{4}{7}\widetilde{p}_{00}\right)\\
\frac{2\tau_{s}}{\tau_{d}}\frac{\widetilde{\beta}}{B}\widetilde{u}_{00}\left(1-\frac{2}{3}\widetilde{p}_{00}\right) & -\frac{\tau_{s}}{\tau_{d}}-\frac{2}{3}\frac{\tau_{s}}{\tau_{d}}\frac{\widetilde{\beta}}{B}\widetilde{u}_{00}^{2} & -\frac{\tau_{s}}{\tau_{d}}\frac{\widetilde{\beta}}{B^{2}}\widetilde{u}_{00}^{2}\left(1-\frac{2}{3}\widetilde{p}_{00}\right)\\
\frac{1}{8}\frac{\tau_{s}}{\tau_{B}}\widetilde{k}\widetilde{u}_{00} & 0 & -\frac{\tau_{s}}{\tau_{B}}
\end{array}\right),
\nonumber 
\end{align}
and
\begin{align}
F_{\widetilde{u}_{00}}\left(\widetilde{u}_{00},p_{00},B\right) \equiv & ~ -\widetilde{u}_{00}+\frac{1}{2}\frac{1}{B}\widetilde{u}_{00}^{2}\left[1-\frac{4}{7}p_{00}\right],\\
F_{p_{00}}\left(\widetilde{u}_{00},p_{00},B\right) \equiv & ~ \frac{\tau_{s}}{\tau_{d}}\left\{ -p_{00}+\frac{1}{B}\widetilde{\beta}\widetilde{u}_{00}^{2}\left[1-\frac{2}{3}p_{00}\right]\right\} ,\\
F_{B}\left(\widetilde{u}_{00},p_{00},B\right) \equiv & ~ \frac{\tau_{s}}{\tau_{B}}\left[-B+1+\frac{1}{16}\widetilde{k}\widetilde{u}_{00}^{2}\right].
\end{align}
By calculating eigenvalues of this matrix for a given fixed point
solution, we can test the stability of each solution. 
In Figure \ref{fig:std_parabola}
the parabolas are plotted from $\widetilde{\beta}=0$ to a point such
that the real part of one of the eigenvalues just turns positive.
The solid line in Figure \ref{fig:std_parabola} maps the parameter
region for the existence of static profiles of $u(\mathbf{x},t)$.

For the stability matrix in Eq. (\ref{eq:trans_std}), since $p_{10}=0$ 
for static states, amplitudal stability and translational stability
are uncoupled.  The $\mathscr{A}_{\rm STD}$ in the $4\times4$ matrix in 
Eq. (\ref{eq:trans_std}) is the same as that given in Eq. (\ref{eq:A_std}).

\subsection{\label{subsec:moving_sol_std} Derivations of Eqs. (\ref{eq:dukkdt}) - (\ref{eq:dBdt})}

To obtain Eqs. (\ref{eq:dukkdt}) - (\ref{eq:dBdt}), we need to deal 
with differentiations
of basis functions $\psi_k\left(\xi\right)$ and 
$\varphi_k\left(\xi\right)$:
\begin{align}
\frac{d\psi_k\left(\xi\right)}{d\xi}
= & ~ \frac{1}{2a}\sqrt{k} \psi_{k-1}\left(\xi\right) - 
\frac{1}{2a}\sqrt{k+1}\psi_{k-1}\left(\xi\right) \\
\frac{d\varphi_k\left(\xi\right)}{d\xi}
= & ~ \frac{1}{\sqrt 2a}\sqrt{k} \varphi_{k-1}\left(\xi\right) - 
\frac{1}{\sqrt 2a}\sqrt{k+1}\varphi_{k-1}\left(\xi\right) 
\end{align}
Combining this result with Eqs. (\ref{eq:uk}), 
(\ref{eq:pk}), we have
\begin{align}
\tau_s \frac{\partial\widetilde{u}\left(\boldsymbol{\xi},t\right)}{\partial t}
 = & ~ \tau_s \sum_{k_0k_1}
\biggl\{\frac{d\widetilde{u}_{k_0k_1}\left(t\right)}{dt}
-\frac{c_0}{2a}\left[\sqrt{k_0+1}\widetilde{u}_{k_0+1,k_1}\left(t\right)
-\sqrt{k_0}\widetilde{u}_{k_0-1,k_1}\left(t\right)\right]\biggr. \nonumber \\
 & ~ \biggl.\frac{c_1}{2a}\left[\sqrt{k_1+1}\widetilde{u}_{k_0,k_1+1}\left(t\right)
-\sqrt{k_1}\widetilde{u}_{k_0,k_1-1}\left(t\right)\right]\biggr\}
\psi_{k_0}\left(\xi_0\right)\psi_{k_1}\left(\xi_1\right) \\
\tau_d \frac{\partial p\left(\boldsymbol{\xi},t\right)}{\partial t}
 = & ~ -\tau_d \sum_{k_0k_1}
\biggl\{\frac{dp_{k_0k_1}\left(t\right)}{dt}
-\frac{c_0}{\sqrt{2}a}\left[\sqrt{k_0+1}p_{k_0+1,k_1}\left(t\right)
-\sqrt{k_0}p_{k_0-1,k_1}\left(t\right)\right]\biggr. \nonumber \\
 & ~ \biggl.\frac{c_1}{\sqrt{2}a}\left[\sqrt{k_1+1}p_{k_0,k_1+1}\left(t\right)
-\sqrt{k_1}p_{k_0,k_1-1}\left(t\right)\right]\biggr\}
\varphi_{k_0}\left(\xi_0\right)\varphi_{k_1}\left(\xi_1\right)
\end{align}
We now obtain the left-hand sides of Eqs. (\ref{eq:dudt}) and (\ref{eq:dpdt}).
For the right-hand sides, by substituting Eqs. (\ref{eq:uk}) and
(\ref{eq:pk}) into right-hand sides of 
Eqs. (\ref{eq:dudt}) and (\ref{eq:dpdt}) and projecting 
them onto corresponding basis functions, we obtain
\begin{align}
 & ~ -\widetilde u\left(\mathbf{x},t\right) + 
\frac{1}{B\left(t\right)}\int d\mathbf{x}
\widetilde J\left(\left|\mathbf{x}-\mathbf{x}^\prime\right|\right)
p\left(\mathbf{x}^\prime,t\right)
\widetilde u\left(\mathbf{x}^\prime,t\right)^2 \nonumber \\
 = & ~
\sum_{k_0k_1} \widetilde u_{k_0k_1}
\psi_{k_0}\left(\xi_0\right)\psi_{k_1}\left(\xi_1\right) \nonumber \\
   & ~ +\frac{1}{B\left(t\right)}\sum_{k_{0}k_{1}}\psi_{k_{0}}\left(\xi_{0}\right)\psi_{k_{1}}\left(\xi_{1}\right)\left(\sum_{n_{0}n_{1}m_{0}m_{1}}C_{n_{0}m_{0}}^{k_{0}}C_{n_{1}m_{1}}^{k_{1}}\widetilde{u}_{n_{0}n_{1}}\widetilde{u}_{m_{0}m_{1}}\right)\nonumber \\
   & ~ -\frac{1}{B\left(t\right)}\sum_{k_{0}k_{1}}\psi_{k_{0}}\left(\xi_{0}\right)\psi_{k_{1}}\left(\xi_{1}\right)\left(\sum_{n_{0}n_{1}m_{0}m_{1}l_{0}l_{1}}D_{n_{0}m_{0}l_{0}}^{k_{0}}D_{n_{1}m_{1}l_{1}}^{k_{1}}\widetilde{u}_{n_{0}n_{1}}\widetilde{u}_{m_{0}m_{1}}p_{l_{0}l_{1}}\right), \nonumber \\ \\
   & ~ 1-p\left(\mathbf{x},t\right)-\tau_{d}\beta p\left(\mathbf{x},t\right)\frac{\widetilde u\left(\mathbf{x},t\right)^{2}}{B\left(t\right)}\nonumber \\
  = & ~ \sum_{k_{0}k_{1}}p_{k_{0}k_{1}}\left(t\right)\varphi_{k_{0}}\left(\xi_{0}\right)\varphi_{k_{1}}\left(\xi_{1}\right)\nonumber \\
   & ~ -\frac{\tau_{d}\beta}{B\left(t\right)}\sum_{k_{0}k_{1}}\varphi_{k_{0}}\left(\xi_{0}\right)\varphi_{k_{1}}\left(\xi_{1}\right)\left(\sum_{n_{0}n_{1}m_{0}m_{1}}E_{n_{0}m_{0}}^{k_{0}}E_{n_{1}m_{1}}^{k_{1}}\widetilde{u}_{n_{0}n_{1}}\widetilde{u}_{m_{0}m_{1}}\right)\nonumber \\
   & ~ +\frac{\tau_{d}\beta}{B\left(t\right)}\sum_{k_{0}k_{1}}\varphi_{k_{0}}\left(\xi_{0}\right)\varphi_{k_{1}}\left(\xi_{1}\right)\left(\sum_{n_{0}n_{1}m_{0}m_{1}}F_{n_{0}m_{0}l_{0}}^{k_{0}}F_{n_{1}m_{1}l_{1}}^{k_{1}}\widetilde{u}_{n_{0}n_{1}}\widetilde{u}_{m_{0}m_{1}}p_{l_{0}l_{1}}\right).
\end{align}
Here argument $t$ of $\widetilde u_{k_0k_1}$ and $p_{k_0k_1}$ are
omitted.  
$\widetilde J\left(\left|\mathbf{x}-\mathbf{x}^\prime\right|\right)
\equiv J_0^{-1}J\left(\left|\mathbf{x}-\mathbf{x}^\prime\right|\right)$. 
Combining expressions of the left-hand and right-hand sides of 
Eqs. (\ref{eq:dudt}) and (\ref{eq:dpdt}), we should obtain 
Eqs. (\ref{eq:dukkdt}) and (\ref{eq:dpkkdt}).
Eq. (\ref{eq:dBdt}) can be obtained easily by substituting
Eq. (\ref{eq:uk}) into Eq. (\ref{dBdt}) and making use
of the orthogonality of $\psi_{k_i}\left(\xi_i\right)$'s.

\subsection{\label{subsec:moving_std} Stability of Moving Bumps on 2D CANNs with STD}

To study the stability issue, let us define
\begin{align}
F\left(\widetilde{u}_{00},\cdots,p_{00},\cdots,B|\widetilde{u}_{k_{0}k_{1}}\right)  \equiv &~ -\widetilde{u}_{k_{0}k_{1}}
+\frac{\tau_{s}c_0}{2a}\left[\sqrt{k_{0}+1}\widetilde{u}_{k_{0}+1,k_{1}}-\sqrt{k_{0}}\widetilde{u}_{k_{0}-1,k_{1}}\right]\nonumber \\
 &~ +\frac{\tau_{s}c_1}{2a}\left[\sqrt{k_{1}+1}\widetilde{u}_{k_{0},k_{1}+1}-\sqrt{k_{1}}\widetilde{u}_{k_{0},k_{1}-1}\right]\nonumber \\
 &~ +\frac{1}{B}\sum_{n_{0}n_{1}m_{0}m_{1}}C_{n_{0}m_{0}}^{k_{0}}C_{n_{1}m_{1}}^{k_{1}}\widetilde{u}_{n_{0}n_{1}}\widetilde{u}_{m_{0}m_{1}}\nonumber \\
 &~ -\frac{1}{B}\sum_{n_{0}n_{1}m_{0}m_{1}l_{0}l_{1}}D_{n_{0}m_{0}l_{0}}^{k_{0}}D_{n_{1}m_{1}l_{1}}^{k_{1}}\widetilde{u}_{n_{0}n_{1}}\widetilde{u}_{m_{0}m_{1}}p_{l_{0}l_{1}}, \nonumber \\ \\
F\left(\widetilde{u}_{00},\cdots,p_{00},\cdots,B|p_{k_{0}k_{1}}\right) \equiv &~ \frac{\tau_s}{\tau_d} \Biggl\{ -p_{k_{0}k_{1}} \Biggr.\nonumber \\
 &~+\frac{\tau_{d}c_0}{\sqrt{2}a}\left[\sqrt{k_{0}+1}p_{k_{0}+1,k_{1}}-\sqrt{k_{0}}p_{k_{0}-1,k_{1}}\right]\nonumber \\
 &~+\frac{\tau_{d}c_1}{\sqrt{2}a}\left[\sqrt{k_{1}+1}p_{k_{0},k_{1}+1}-\sqrt{k_{1}}p_{k_{0},k_{1}-1}\right]\nonumber \\
 &~ +\frac{\widetilde{\beta}}{B}\sum_{n_{0}n_{1}m_{0}m_{1}}E_{n_{0}m_{0}}^{k_{0}}E_{n_{1}m_{1}}^{k_{1}}\widetilde{u}_{n_{0}n_{1}}\widetilde{u}_{m_{0}m_{1}}\nonumber \\
 &~ \Biggl.-\frac{\widetilde{\beta}}{B}\sum_{n_{0}n_{1}m_{0}m_{1}l_{0}l_{1}}F_{n_{0}m_{0}l_{0}}^{k_{0}}F_{n_{1}m_{1}l_{1}}^{k_{1}}\widetilde{u}_{n_{0}n_{1}}\widetilde{u}_{m_{0}m_{1}}p_{l_{0}l_{1}}\Biggr\}, \nonumber \\ \\
F\left(\widetilde{u}_{00},\cdots,p_{00},\cdots,B|B\right) \equiv &~ 
\frac{\tau_s}{\tau_B}\Biggl(-B+1+\frac{1}{32\pi a^{2}}\widetilde{k}\sum_{k_{0}k_{1}}\widetilde{u}_{k_{0}k_{1}}^{2}\Biggr).
\end{align}
The stability of fixed point solutions can be determined by 
considering the stability matrix given by
\begin{align}
 & ~ \mathscr{A}_{\rm STD}^{\rm moving} \nonumber \\
 \equiv & ~
\left(\begin{array}{ccccc}
\frac{\partial F\left(\widetilde{u}_{00},\cdots,p_{00},\cdots,B|\widetilde{u}_{00}\right)}{\partial\widetilde{u}_{00}} & \cdots & \frac{\partial F\left(\widetilde{u}_{00},\cdots,p_{00},\cdots,B|\widetilde{u}_{00}\right)}{\partial p_{00}} & \cdots & \frac{\partial F\left(\widetilde{u}_{00},\cdots,p_{00},\cdots,B|\widetilde{u}_{00}\right)}{\partial B}\\
\vdots & \ddots & \vdots & \ddots & \vdots\\
\frac{\partial F\left(\widetilde{u}_{00},\cdots,p_{00},\cdots,B|p_{00}\right)}{\partial\widetilde{u}_{00}} & \cdots & \frac{\partial F\left(\widetilde{u}_{00},\cdots,p_{00},\cdots,B|p_{00}\right)}{\partial p_{00}} & \cdots & \frac{\partial F\left(\widetilde{u}_{00},\cdots,p_{00},\cdots,B|p_{00}\right)}{\partial B}\\
\vdots & \ddots & \vdots & \ddots & \vdots\\
\frac{\partial F\left(\widetilde{u}_{00},\cdots,p_{00},\cdots,B|B\right)}{\partial\widetilde{u}_{00}} & \cdots & \frac{\partial F\left(\widetilde{u}_{00},\cdots,p_{00},\cdots,B|B\right)}{\partial p_{00}} & \cdots & \frac{\partial F\left(\widetilde{u}_{00},\cdots,p_{00},\cdots,B|B\right)}{\partial B}
\end{array}\right). \nonumber \\
\end{align}
For dynamical variables near their fixed point solutions, we let
\begin{align}
\widetilde{u}_{k_{0}k_{1}}\left(t\right) \equiv & ~ \widetilde{u}_{k_{0}k_{1}}^{*}+\delta\widetilde{u}_{k_{0}k_{1}}\left(t\right),\\
p_{k_{0}k_{1}}\left(t\right) \equiv & ~ p_{k_{0}k_{1}}^{*}+\delta p_{k_{0}k_{1}}\left(t\right),\\
B\left(t\right) \equiv & ~ B^{*}+\delta B\left(t\right),
\end{align}
where $\widetilde{u}_{k_{0}k_{1}}^{*}$, $p_{k_{0}k_{1}}^{*}$ and $B^{*}$ is 
the fixed point solution. Then the dynamics of $\delta\widetilde{u}_{k_{0}k_{1}}\left(t\right)$, $\delta p_{k_{0}k_{1}}\left(t\right)$ and $\delta B\left(t\right)$ can be formulated as 
\begin{equation}
\tau_s \frac{d}{dt}
\left(\begin{array}{c}
\delta\widetilde{u}_{00}\left(t\right) \\
\vdots \\
\delta p_{00}\left(t\right) \\
\vdots \\
\delta B\left(t\right)
\end{array}\right)
=
\mathscr{A}_{\rm STD}^{\rm moving}
\left(\begin{array}{c}
\delta\widetilde{u}_{00}\left(t\right) \\
\vdots \\
\delta p_{00}\left(t\right) \\
\vdots \\
\delta B\left(t\right)
\end{array}\right).
\end{equation}
The fixed point solution is stable only if the maximum of the
real parts of eigenvalues of \\$
\left.\mathscr{A}_{\rm STD}^{\rm moving}\right|_{
\widetilde u_{00}^{*},\dots,p_{00}^{*},\dots,B^*
}$ is non-positive. In general, those eigenvalues can only be 
calculated by numerical methods. The predicted phase boundary
separating the moving and silent phases shown in 
Figure \ref{fig:phase_diagram_std} can be deduced, and the prediction
can be verified by simulations.

\subsection{\label{subsec:static_sub} Amplitudal Stability of Bumps on 2D CANNs with SFA}

For our convenience we define
\begin{align}
K_{\widetilde{u}_{00}}\left(\widetilde u_{00},\widetilde v_{00}, B\right)
\equiv & ~ -\widetilde u_{00} + \frac{1}{2} 
\frac{\widetilde u_{00}^2}{B} - \widetilde{v}_{00},\\
K_{\widetilde{v}_{00}}\left(\widetilde u_{00},\widetilde v_{00}, B\right)
\equiv & ~ \frac{\tau_s}{\tau_i}\left(-\widetilde{v}_{00} + \gamma \widetilde{u}_{00}\right), \\
K_{B}\left(\widetilde u_{00},\widetilde v_{00}, B\right)
\equiv & ~ \frac{\tau_s}{\tau_i}\left(-B + 1 + \frac{1}{16}\widetilde k \widetilde u_{00}^2\right).
\end{align}
Then the stability matrix is given by
\begin{align}
\mathscr{A}_{\rm SFA} = & ~
\left(\begin{array}{ccc}
\frac{\partial K_{\widetilde{u}_{00}}\left(\widetilde u_{00},\widetilde v_{00}, B\right)}{\partial \widetilde u_{00}} &
\frac{\partial K_{\widetilde{u}_{00}}\left(\widetilde u_{00},\widetilde v_{00}, B\right)}{\partial \widetilde v_{00}} &
\frac{\partial K_{\widetilde{u}_{00}}\left(\widetilde u_{00},\widetilde v_{00}, B\right)}{\partial B} \\
\frac{\partial K_{\widetilde{v}_{00}}\left(\widetilde u_{00},\widetilde v_{00}, B\right)}{\partial \widetilde u_{00}} &
\frac{\partial K_{\widetilde{v}_{00}}\left(\widetilde u_{00},\widetilde v_{00}, B\right)}{\partial \widetilde v_{00}} &
\frac{\partial K_{\widetilde{v}_{00}}\left(\widetilde u_{00},\widetilde v_{00}, B\right)}{\partial B} \\
\frac{\partial K_{B}\left(\widetilde u_{00},\widetilde v_{00}, B\right)}{\partial \widetilde u_{00}} &
\frac{\partial K_{B}\left(\widetilde u_{00},\widetilde v_{00}, B\right)}{\partial \widetilde v_{00}} &
\frac{\partial K_{B}\left(\widetilde u_{00},\widetilde v_{00}, B\right)}{\partial B} 
\end{array}\right) \nonumber \\
= & ~
\left(\begin{array}{ccc}
-1+2\left(1+\gamma\right) & -1 & -2\left(1+\gamma\right)^2 \\
\frac{\tau_s}{\tau_i}\gamma & -\frac{\tau_s}{\tau_i} & 0 \\
\frac{\tau_s}{\tau_B}
\frac{1\pm\sqrt{1-\left(1+\gamma\right)^2\widetilde k}}{2\left(1+\gamma\right)} & 0 & -\frac{\tau_s}{\tau_B}
\end{array}\right)
\end{align}
The dynamics of distortions becomes
\begin{equation}
\tau_s
\frac{d}{dt}
\left(\begin{array}{c}
\delta\widetilde u_{00}\left(t\right) \\
\delta\widetilde v_{00}\left(t\right) \\
\delta B \left(t\right) 
\end{array}\right)
=
\mathscr{A}_{\rm SFA}
\left(\begin{array}{c}
\delta\widetilde u_{00}\left(t\right) \\
\delta\widetilde v_{00}\left(t\right) \\
\delta B \left(t\right) 
\end{array}\right)
\end{equation}
If real parts of all eigenvalues of $\mathscr{A}_{\rm SFA}$ are 
non-positive, the static fixed point solution is stable.  For the 
$\mathscr{A}_{\rm SFA}$ in Eq. (\ref{eq:trans_stable_sub}), since
symmetric terms and asymmetric terms are uncoupled for static bumps, 
that $\mathscr{A}_{\rm SFA}$ is the same as the matrix we have shown in 
this section.

\subsection{\label{subsec:moving_Cknm_sub} Recurrence Relation of $C^k_{nm}$ for CANNs with SFA}

We have defined $C_{nm}^k$ in Eq. (\ref{eq:Cknm_sub}). 
\begin{align}
C_{nm}^{k} \equiv & ~ \frac{1}{\sqrt{2\pi}a}\int dx\psi_{k}\left(x\right)\int dx'\exp\left[-\frac{\left(x-x'\right)^{2}}{2a^{2}}\right]\psi_{n}\left(x'\right)\psi_{m}\left(x'\right)\\
 = & ~ \frac{1}{\sqrt{2\pi}a}A_{k}A_{n}A_{m}\int dx\int dx'H_{k}\left(\frac{x}{a}\right)\exp\left[-\frac{x^{2}}{4a^{2}}\right] \nonumber \\
  & ~ \qquad\qquad\qquad\exp\left[-\frac{\left(x-x'\right)^{2}}{2a^{2}}\right]H_{n}\left(\frac{x'}{a}\right)H_{m}\left(\frac{x'}{a}\right)\exp\left[-\frac{x'^{2}}{2a^{2}}\right], \nonumber \\
\end{align}
where 
\begin{equation}
A_{n}=\frac{1}{\sqrt{\sqrt{2\pi}an!}}.
\end{equation}

To derive the recurrence relations, we first consider
\begin{align}
  &~ \int dxH_{k}\left(\frac{x}{a}\right)\exp\left[-\frac{x^{2}}{4a^{2}}\right]\exp\left[-\frac{\left(x-x'\right)^{2}}{2a^{2}}\right]\nonumber\\
  = &~ \left(-1\right)^{k}\int dx\exp\left[-\frac{\left(x-x'\right)^{2}}{2a^{2}}\right]\exp\left(\frac{x^{2}}{4a^{2}}\right)\frac{d^{k}}{d\left(\frac{x}{a}\right)^{k}}\exp\left(-\frac{x^{2}}{2a^{2}}\right)\nonumber\\
  = &~ \left(-1\right)^{k}a^{k}\int d\left[\frac{d^{k-1}}{dx^{k-1}}\exp\left(-\frac{x^{2}}{2a^{2}}\right)\right]\exp\left[-\frac{\left(x-x'\right)^{2}}{2a^{2}}\right]\exp\left(\frac{x^{2}}{4a^{2}}\right) \nonumber\\
  = &~ \left(-1\right)^{k-1}a^{k}\int d\left\{ \exp\left[-\frac{\left(x-x'\right)^{2}}{2a^{2}}+\frac{x^{2}}{4a^{2}}\right]\right\} \frac{d^{k-1}}{dx^{k-1}}\exp\left(-\frac{x^{2}}{2a^{2}}\right)\nonumber\\
  = &~ \left(-1\right)^{k-1}a^{k}\int d\left\{ \exp\left[-\frac{1}{4a^{2}}\left(2x^{2}-4xx'+2x'^{2}-x^{2}\right)\right]\right\} \frac{d^{k-1}}{dx^{k-1}}\exp\left(-\frac{x^{2}}{2a^{2}}\right)\nonumber\\
  = &~ \left(-1\right)^{k-1}a^{k}\exp\left[\frac{x'^{2}}{2a^{2}}\right]\int d\left\{ \exp\left[-\frac{\left(x-2x'\right)^{2}}{4a^{2}}\right]\right\} \frac{d^{k-1}}{dx^{k-1}}\exp\left(-\frac{x^{2}}{2a^{2}}\right)\nonumber\\
  = &~ \left(-1\right)^{k-1}a^{k}\exp\left[\frac{x'^{2}}{2a^{2}}\right]\int dx\frac{d}{dx}\exp\left[-\frac{\left(x-2x'\right)^{2}}{4a^{2}}\right]\frac{d^{k-1}}{dx^{k-1}}\exp\left(-\frac{x^{2}}{2a^{2}}\right)\nonumber\\
  = &~ \left(-1\right)^{k}a^{k}\frac{1}{2}\exp\left[\frac{x'^{2}}{2a^{2}}\right]\frac{d}{dx'}\int dx\exp\left[-\frac{\left(x-2x'\right)^{2}}{4a^{2}}\right]\frac{d^{k-1}}{dx^{k-1}}\exp\left(-\frac{x^{2}}{2a^{2}}\right)\nonumber \\
  = &~ \left(-1\right)^{k}a^{k}\frac{1}{2}\exp\left[\frac{x'^{2}}{2a^{2}}\right]\frac{d}{dx'}\int d\left[\frac{d^{k-2}}{dx^{k-2}}\exp\left(-\frac{x^{2}}{2a^{2}}\right)\right]\exp\left[-\frac{\left(x-2x'\right)^{2}}{4a^{2}}\right]\nonumber\\
  = &~ \left(-1\right)^{k-1}a^{k}\frac{1}{2}\exp\left[\frac{x'^{2}}{2a^{2}}\right]\frac{d}{dx'}\int d\left\{ \exp\left[-\frac{\left(x-2x'\right)^{2}}{4a^{2}}\right]\right\} \frac{d^{k-2}}{dx^{k-2}}\exp\left(-\frac{x^{2}}{2a^{2}}\right)\nonumber\\
  = &~ \left(-1\right)^{k}a^{k}\frac{1}{2^{2}}\exp\left[\frac{x'^{2}}{2a^{2}}\right]\frac{d^{2}}{dx'^{2}}\int d\exp\left[-\frac{\left(x-2x'\right)^{2}}{4a^{2}}\right]\frac{d^{k-2}}{dx^{k-2}}\exp\left(-\frac{x^{2}}{2a^{2}}\right)\nonumber\\
   &~ \vdots\nonumber\\
  = &~ \left(-1\right)^{k}a^{k}\frac{1}{2^{k}}\exp\left[\frac{x'^{2}}{2a^{2}}\right]\frac{d^{k}}{dx'^{k}}\int d\exp\left[-\frac{\left(x-2x'\right)^{2}}{4a^{2}}\right]\exp\left(-\frac{x^{2}}{2a^{2}}\right)\nonumber\\
  = &~ \left(-1\right)^{k}a^{k}\frac{1}{2^{k}}\sqrt{\frac{4\pi}{3}}a\exp\left[\frac{x'^{2}}{2a^{2}}\right]\frac{d^{k}}{dx'^{k}}\exp\left[-\frac{2x'^{2}}{3a^{2}}\right]\nonumber\\
  = &~ \left(-1\right)^{k}a^{k}\frac{1}{2^{k}}\sqrt{\frac{4\pi}{3}}a\exp\left[\frac{x'^{2}}{2a^{2}}\right]\frac{d^{k}}{dx'^{k}}\exp\left[-\frac{1}{2}\left(\frac{2x}{\sqrt{3}a}\right)^{2}\right]\nonumber\\
  = &~ \left(-1\right)^{k}a^{k}\frac{1}{2^{k}}\sqrt{\frac{4\pi}{3}}a\left(\frac{2}{\sqrt{3}a}\right)^{k}\exp\left[\frac{x'^{2}}{2a^{2}}-\frac{1}{2}\left(\frac{2x'}{\sqrt{3}a}\right)^{2}\right]\nonumber\\
   &~ \qquad\qquad\qquad\exp\left[\frac{1}{2}\left(\frac{2x'}{\sqrt{3}a}\right)^{2}\right]\frac{d^{k}}{d\left(\frac{2x'}{\sqrt{3}a}\right)^{k}}\exp\left[-\frac{1}{2}\left(\frac{2x'}{\sqrt{3}a}\right)^{2}\right]\nonumber\\
  = &~ \frac{2\sqrt{\pi}a}{3^{\frac{k+1}{2}}}\exp\left(-\frac{x'^{2}}{6a^{2}}\right)H_{k}\left(\frac{2x'}{\sqrt{3}a}\right).
\end{align}
Here we obtain this identity by using multiple integration by parts. Then, we have
\begin{align}
{C}_{nm}^{k} = &~ \frac{\sqrt{2}}{3^{\frac{k+1}{2}}}A_{k}A_{n}A_{m}\int dx'\exp\left(-\frac{2x'^{2}}{3a^{2}}\right)H_{k}\left(\frac{2x'}{\sqrt{3}a}\right)H_{n}\left(\frac{x'}{a}\right)H_{m}\left(\frac{x'}{a}\right)\nonumber \\
 = &~ \frac{\sqrt{2}}{3^{\frac{k+1}{2}}}\frac{1}{\left(2\pi\right)^{\frac{3}{4}}a^{\frac{3}{2}}\sqrt{k!n!m!}}\int dx\exp\left(-\frac{2x^{2}}{3a^{2}}\right)H_{k}\left(\frac{2x}{\sqrt{3}a}\right)H_{n}\left(\frac{x}{a}\right)H_{m}\left(\frac{x}{a}\right). \nonumber \\
\end{align}
To relate $C_{nm}^k$ and $C_{n^\prime m^\prime}^{k^\prime}$ we need to use the recursion
relations of probabilist's Hermite polynomials:
\begin{align}
H_{n+1}\left(x\right) = & ~ xH_{n}\left(x\right)-H_{n}'\left(x\right)\text{ and }\\
H_{n}'\left(x\right) = & ~ nH_{n-1}\left(x\right).
\end{align}
Then we can derive
\begin{align}
{C}_{nm}^{k}  = &~ \frac{\sqrt{2}}{3^{\frac{k+1}{2}}}\frac{1}{\left(2\pi\right)^{\frac{3}{4}}a^{\frac{3}{2}}\sqrt{k!n!m!}}\int dx\exp\left(-\frac{2x^{2}}{3a^{2}}\right)H_{k}\left(\frac{2x}{\sqrt{3}a}\right)H_{n}\left(\frac{x}{a}\right)H_{m}\left(\frac{x}{a}\right)\nonumber \\
  = &~ \frac{\sqrt{2}}{3^{\frac{k+1}{2}}}\frac{1}{\left(2\pi\right)^{\frac{3}{4}}a^{\frac{3}{2}}\sqrt{k!n!m!}}\nonumber \\
   &~ \quad\int dx\exp\left(-\frac{2x^{2}}{3a^{2}}\right)\left[\frac{2x}{\sqrt{3}a}H_{k-1}\left(\frac{2x}{\sqrt{3}a}\right)-\left(k-1\right)H_{k-2}\left(\frac{2x}{\sqrt{3}a}\right)\right] \nonumber \\
  &~ \quad \quad \quad\quad \quad \quad\quad \quad \quad H_{n}\left(\frac{x}{a}\right)H_{m}\left(\frac{x}{a}\right)\nonumber \\
  = &~ \frac{\sqrt{2}}{3^{\frac{k+1}{2}}}\frac{1}{\left(2\pi\right)^{\frac{3}{4}}a^{\frac{3}{2}}\sqrt{k!n!m!}}\int dx\exp\left(-\frac{2x^{2}}{3a^{2}}\right)\frac{2x}{\sqrt{3}a}\nonumber \\
 &~ \quad \quad \quad\quad \quad \quad\quad \quad \quad \quad \quad \quad H_{k-1}\left(\frac{2x}{\sqrt{3}a}\right)H_{n}\left(\frac{x}{a}\right)H_{m}\left(\frac{x}{a}\right)\nonumber \\
   &~ -\frac{\sqrt{2}}{3^{\frac{k+1}{2}}}\frac{1}{\left(2\pi\right)^{\frac{3}{4}}a^{\frac{3}{2}}\sqrt{k!n!m!}}\int dx\exp\left(-\frac{2x^{2}}{3a^{2}}\right)\left(k-1\right) \nonumber \\
  &~\quad \quad \quad\quad \quad \quad\quad \quad \quad\quad \quad \quad H_{k-2}\left(\frac{2x}{\sqrt{3}a}\right)H_{n}\left(\frac{x}{a}\right)H_{m}\left(\frac{x}{a}\right)\nonumber \\
  = &~ -\frac{1}{3}\sqrt{\frac{k-1}{k}}{C}_{nm}^{k-2}+\frac{\sqrt{2}}{3^{\frac{k+1}{2}}}\frac{1}{\left(2\pi\right)^{\frac{3}{4}}a^{\frac{3}{2}}\sqrt{k!n!m!}}\left(-\frac{\sqrt{3}}{2}a\right)\nonumber \\
   &~ \quad\int d\left[\exp\left(-\frac{2x^{2}}{3a^{2}}\right)\right]H_{k-1}\left(\frac{2x}{\sqrt{3}a}\right)H_{n}\left(\frac{x}{a}\right)H_{m}\left(\frac{x}{a}\right)\nonumber \\
  = &~ -\frac{1}{3}\sqrt{\frac{k-1}{k}}{C}_{nm}^{k-2}+\frac{\sqrt{2}}{3^{\frac{k+1}{2}}}\frac{1}{\left(2\pi\right)^{\frac{3}{4}}a^{\frac{3}{2}}\sqrt{k!n!m!}}\frac{\sqrt{3}}{2}a\nonumber \\
   &~ \quad\int dx\exp\left(-\frac{2x^{2}}{3a^{2}}\right)\left[H_{k-1}'\left(\frac{2x}{\sqrt{3}a}\right)H_{n}\left(\frac{x}{a}\right)H_{m}\left(\frac{x}{a}\right)\frac{2}{\sqrt{3}a}\right.\nonumber \\
   &~ \quad \quad \quad\quad \quad \quad\quad \quad \quad+H_{k-1}\left(\frac{2x}{\sqrt{3}a}\right)H_{n}'\left(\frac{x}{a}\right)H_{m}\left(\frac{x}{a}\right)\frac{1}{a}\nonumber\\
  &~ \left.\quad \quad \quad\quad \quad\quad \quad \quad \quad +H_{k-1}\left(\frac{2x}{\sqrt{3}a}\right)H_{n}\left(\frac{x}{a}\right)H_{m}'\left(\frac{x}{a}\right)\frac{1}{a}\right]\nonumber \\
  = &~ -\frac{1}{3}\sqrt{\frac{k-1}{k}}{C}_{nm}^{k-2}+\frac{1}{3}\frac{\sqrt{3}}{2}a\frac{2}{\sqrt{3}a}\frac{k-1}{\sqrt{k\left(k-1\right)}}{C}_{nm}^{k-2}\nonumber \\
   &~ \quad \quad \quad +\frac{1}{\sqrt{3}}\frac{\sqrt{3}}{2}a\frac{1}{a}\frac{n}{\sqrt{kn}}{C}_{n-1,m}^{k-1}+\frac{1}{\sqrt{3}}\frac{\sqrt{3}}{2}a\frac{1}{a}\frac{m}{\sqrt{km}}{C}_{n,m-1}^{k-1}\nonumber \\
  = &~ \frac{1}{2}\sqrt{\frac{n}{k}}{C}_{n-1,m}^{k-1}+\frac{1}{2}\sqrt{\frac{m}{k}}{C}_{n,m-1}^{k-1}\label{eq:C_nmk_k_iter}
\end{align}
Also,
\begin{align}
{C}_{nm}^{k}  = &~ \frac{\sqrt{2}}{3^{\frac{k+1}{2}}}\frac{1}{\left(2\pi\right)^{\frac{3}{4}}a^{\frac{3}{2}}\sqrt{k!n!m!}}\nonumber \\
   &~ \quad\int dx\exp\left(-\frac{2x^{2}}{3a^{2}}\right)H_{k}\left(\frac{2x}{\sqrt{3}a}\right)\left[\frac{x}{a}H_{n-1}\left(\frac{x}{a}\right)-H_{n-1}'\left(\frac{x}{a}\right)\right]H_{m}\left(\frac{x}{a}\right)\nonumber \\
  = &~ \frac{\sqrt{2}}{3^{\frac{k+1}{2}}}\frac{1}{\left(2\pi\right)^{\frac{3}{4}}a^{\frac{3}{2}}\sqrt{k!n!m!}}\int dx\exp\left(-\frac{2x^{2}}{3a^{2}}\right)H_{k}\left(\frac{2x}{\sqrt{3}a}\right)\nonumber \\
  &~ \quad \quad \quad \quad \quad \quad \quad \quad \quad \quad \left[\frac{x}{a}H_{n-1}\left(\frac{x}{a}\right)-\left(n-1\right)H_{n-2}\left(\frac{x}{a}\right)\right]H_{m}\left(\frac{x}{a}\right)\nonumber \\
  = &~ \frac{\sqrt{2}}{3^{\frac{k+1}{2}}}\frac{1}{\left(2\pi\right)^{\frac{3}{4}}a^{\frac{3}{2}}\sqrt{k!n!m!}}\int dx\frac{x}{a}\exp\left(-\frac{2x^{2}}{3a^{2}}\right)\nonumber \\
  &~ \quad \quad \quad \quad \quad \quad \quad \quad \quad \quad \quad \quad H_{k}\left(\frac{2x}{\sqrt{3}a}\right) H_{n-1}\left(\frac{x}{a}\right)H_{m}\left(\frac{x}{a}\right)\nonumber \\
   &~ -\frac{\sqrt{2}}{3^{\frac{k+1}{2}}}\frac{1}{\left(2\pi\right)^{\frac{3}{4}}a^{\frac{3}{2}}\sqrt{k!n!m!}}\left(n-1\right)\int dx\exp\left(-\frac{2x^{2}}{3a^{2}}\right) \nonumber \\
  &~  \quad \quad \quad \quad \quad \quad \quad \quad \quad \quad \quad \quad \quad \quad \quad \quad H_{k}\left(\frac{2x}{\sqrt{3}a}\right)H_{n-2}\left(\frac{x}{a}\right)H_{m}\left(\frac{x}{a}\right)\nonumber \\
  = &~ -\sqrt{\frac{n-1}{n}}{C}_{n-2,m}^{k}+\frac{\sqrt{2}}{3^{\frac{k+1}{2}}}\frac{1}{\left(2\pi\right)^{\frac{3}{4}}a^{\frac{3}{2}}\sqrt{k!n!m!}}\left(-\frac{3a}{4}\right)\nonumber \\
   &~ \quad\int d\left[\exp\left(-\frac{2x^{2}}{3a^{2}}\right)\right]H_{k}\left(\frac{2x}{\sqrt{3}a}\right)H_{n-1}\left(\frac{x}{a}\right)H_{m}\left(\frac{x}{a}\right)\nonumber \\
  = &~ -\sqrt{\frac{n-1}{n}}{C}_{n-2,m}^{k}+\frac{\sqrt{2}}{3^{\frac{k+1}{2}}}\frac{1}{\left(2\pi\right)^{\frac{3}{4}}a^{\frac{3}{2}}\sqrt{k!n!m!}}\left(\frac{3a}{4}\right)\nonumber \\
   &~ \quad\int dx\exp\left(-\frac{2x^{2}}{3a^{2}}\right)\left[H_{k}'\left(\frac{2x}{\sqrt{3}a}\right)H_{n-1}\left(\frac{x}{a}\right)H_{m}\left(\frac{x}{a}\right)\frac{2}{\sqrt{3}a}\right.\nonumber \\
   &~\quad \quad \quad \quad \quad \quad \quad \quad \quad \quad +H_{k}\left(\frac{2x}{\sqrt{3}a}\right)H_{n-1}'\left(\frac{x}{a}\right)H_{m}\left(\frac{x}{a}\right)\frac{1}{a} \nonumber \\
  &~ \quad \quad \quad \quad\quad \quad \quad \quad \quad \quad \left. +H_{k}\left(\frac{2x}{\sqrt{3}a}\right)H_{n-1}\left(\frac{x}{a}\right)H_{m}'\left(\frac{x}{a}\right)\frac{1}{a}\right] \nonumber \\
  = &~ -\sqrt{\frac{n-1}{n}}{C}_{n-2,m}^{k}+\frac{1}{\sqrt{3}}\frac{3a}{4}\frac{2}{\sqrt{3}a}\frac{k}{\sqrt{kn}}{C}_{n-1,m}^{k-1}\nonumber \\
   &~ \qquad+\frac{3a}{4}\frac{1}{a}\frac{n-1}{\sqrt{n\left(n-1\right)}}{C}_{n-2,m}^{k}+\frac{3a}{4}\frac{1}{a}\frac{m}{\sqrt{nm}}{C}_{n-1,m-1}^{k}\nonumber \\
  = &~ -\frac{1}{4}\sqrt{\frac{n-1}{n}}{C}_{n-2,m}^{k}+\frac{1}{2}\sqrt{\frac{k}{n}}{C}_{n-1,m}^{k-1}+\frac{3}{4}\sqrt{\frac{m}{n}}{C}_{n-1,m-1}^{k}\label{eq:C_nmk_n_iter}\\
{C}_{nm}^{k}  = &~ -\frac{1}{4}\sqrt{\frac{m-1}{m}}{C}_{n,m-2}^{k}+\frac{1}{2}\sqrt{\frac{k}{m}}{C}_{n,m-1}^{k-1}+\frac{3}{4}\sqrt{\frac{n}{m}}{C}_{n-1,m-1}^{k}\label{eq:C_nmk_m_iter}
\end{align}
For $k=m=n=0$, we have 
\begin{align}
{C}_{00}^{0}  = &~ \frac{\sqrt{2}}{\sqrt{3}}\frac{1}{\left(2\pi\right)^{\frac{3}{4}}a^{\frac{3}{2}}}\int dx\exp\left(-\frac{2x^{2}}{3a^{2}}\right)\nonumber \\
 = &~ \frac{\sqrt{2}}{\sqrt{3}}\frac{1}{\left(2\pi\right)^{\frac{3}{4}}a^{\frac{3}{2}}}\sqrt{\frac{3\pi a^{2}}{2}}\nonumber \\
 = &~ \frac{1}{\sqrt{\sqrt{8\pi}a}}
\end{align}
By using Eqs. (\ref{eq:C_nmk_k_iter}), (\ref{eq:C_nmk_n_iter}) and 
(\ref{eq:C_nmk_m_iter}) and $C_{00}^0$, we can generate any
$C_{nm}^k$ needed for numerical computations.

\subsection{\label{subsec:moving_speed_sub} Intrinsic Speed of Moving Bumps on 2D CANNs with SFA}

Here we consider terms up to $k_0 + k_1 = 2$.  And for simplicity
we assume 
$\mathbf{c} = \left(c_0,c_1\right) = \left(c,0\right)$. Then,
\begin{align}
u\left(\boldsymbol{\xi},t\right)  \approx &~ \sum_{k_{0}+k_{1}\le2}u_{k_{0}k_{1}}\left(t\right)\psi_{k_{0}}\left(\xi_{0}\right)\psi_{k_{1}}\left(\xi_{1}\right) \nonumber \\
  = &~ u_{00}\left(t\right)\psi_{0}\left(\xi_{0}\right)\psi_{0}\left(\xi_{1}\right) \nonumber \\
    &~ +u_{02}\left(t\right)\psi_{0}\left(\xi_{0}\right)\psi_{2}\left(\xi_{1}\right)+u_{20}\left(t\right)\psi_{2}\left(\xi_{0}\right)\psi_{0}\left(\xi_{1}\right),\\
v\left(\boldsymbol{\xi},t\right)  = &~ v_{00}\left(t\right)\psi_{0}\left(\xi_{0}\right)\psi_{0}\left(\xi_{1}\right)+v_{10}\left(t\right)\psi_{1}\left(\xi_{0}\right)\psi_{0}\left(\xi_{1}\right) \nonumber \\
   &~ +v_{02}\left(t\right)\psi_{0}\left(\xi_{0}\right)\psi_{2}\left(\xi_{1}\right)+v_{20}\left(t\right)\psi_{2}\left(\xi_{0}\right)\psi_{0}\left(\xi_{1}\right).
\end{align}
Note that, since $c_1=0$, terms that are asymmetric along
the $\xi_1$-direction are dropped.
The derived speed should be applicable to
any direction because the model is homogeneous.

By projection and orthogonality of the basis functions, 
Eqs. (\ref{eq:dudt}) and (\ref{eq:dvdt_sub}) give
\begin{align}
0 = &~ -\widetilde{u}_{00}-\widetilde{v}_{00}\nonumber \\
   &~ +\frac{1}{B}\left(C_{00}^{0}\right)^{2}\widetilde{u}_{00}^{2}+\frac{1}{B}C_{00}^{0}C_{22}^{0}\widetilde{u}_{02}^{2}+\frac{1}{B}C_{22}^{0}C_{00}^{0}\widetilde{u}_{20}^{2}\nonumber \\
   &~ +2\frac{1}{B}C_{00}^{0}C_{02}^{0}\widetilde{u}_{00}\widetilde{u}_{02}+2\frac{1}{B}C_{02}^{0}C_{00}^{0}\widetilde{u}_{00}\widetilde{u}_{20}+2\frac{1}{B}\left(C_{02}^{0}\right)^{2}\widetilde{u}_{02}\widetilde{u}_{20}\\
0  = &~ \widetilde{v}_{10}+\frac{\tau_{s}c}{2a}\left(\sqrt{1}\widetilde{u}_{00}-\sqrt{2}\widetilde{u}_{20}\right)\label{eq:App_F_7}\\
0  = &~ -\widetilde{u}_{02}-\widetilde{v}_{02}+\frac{1}{B}C_{00}^{0}C_{00}^{2}\widetilde{u}_{00}^{2}+\frac{1}{B}C_{00}^{0}C_{22}^{2}\widetilde{u}_{02}^{2}+\frac{1}{B}C_{22}^{0}C_{00}^{2}\widetilde{u}_{20}^{2}\nonumber \\
   &~ +2\frac{1}{B}C_{00}^{0}C_{02}^{2}\widetilde{u}_{00}\widetilde{u}_{02}+2\frac{1}{B}C_{02}^{0}C_{00}^{2}\widetilde{u}_{00}\widetilde{u}_{20}\nonumber \\
   &~ +2\frac{1}{B}C_{02}^{0}C_{20}^{2}\widetilde{u}_{02}\widetilde{u}_{20}\\
0  = &~ -\widetilde{u}_{20}-\widetilde{v}_{20}+\frac{1}{B}C_{00}^{2}C_{00}^{0}\widetilde{u}_{00}^{2}+\frac{1}{B}C_{00}^{2}C_{22}^{0}\widetilde{u}_{02}^{2}+\frac{1}{B}C_{22}^{2}C_{00}^{0}\widetilde{u}_{20}^{2}\nonumber \\
   &~ +2\frac{1}{B}C_{00}^{2}C_{02}^{0}\widetilde{u}_{00}\widetilde{u}_{02}+2\frac{1}{B}C_{02}^{2}C_{00}^{0}\widetilde{u}_{00}\widetilde{u}_{20}\nonumber \\
   &~ +2\frac{1}{B}C_{02}^{2}C_{20}^{0}\widetilde{u}_{02}\widetilde{u}_{20}\\
0  = &~ -\widetilde{v}_{00}+\gamma\widetilde{u}_{00}+\frac{\tau_{i}c}{2a}\widetilde{v}_{10} \label{eq:App_F_10}\\
0  = &~ -\widetilde{v}_{10}-\frac{\tau_{i}c}{2a}\left(\sqrt{1}\widetilde{v}_{00}-\sqrt{2}\widetilde{v}_{20}\right) \label{eq:App_F_11}\\
0  = &~ -\widetilde{v}_{02}+\gamma\widetilde{u}_{02} \label{eq:App_F_12}\\
0  = &~ -\widetilde{v}_{20}+\gamma\widetilde{u}_{20}-\frac{\tau_{i}c}{2a}\left(\sqrt{2}\widetilde{v}_{10}\right)\label{eq:App_F_13}
\end{align} 
From these equations we can deduce the ratio between 
$\widetilde{u}_{00}-\sqrt{2}\widetilde{u}_{20}$
and $\widetilde{v}_{00}-\sqrt{2}\widetilde{v}_{20}$ 
\begin{align}
0  = & ~ \widetilde{v}_{10}+\frac{\tau_{s}c}{2a}\left(\widetilde{u}_{00}-\sqrt{2}\widetilde{u}_{20}\right)\nonumber \\
0  = & ~ \widetilde{v}_{10}+\frac{\tau_{i}c}{2a}\left(\widetilde{v}_{00}-\sqrt{2}\widetilde{v}_{20}\right) \nonumber\\
\Rightarrow \widetilde{v}_{00}-\sqrt{2}\widetilde{v}_{20}  = & ~ \frac{\tau_{s}}{\tau_{i}}\left(\widetilde{u}_{00}-\sqrt{2}\widetilde{u}_{20}\right)
\end{align}
Combining Eqs. (\ref{eq:App_F_7}), (\ref{eq:App_F_10}) and (\ref{eq:App_F_13}), we have
\begin{align}
\widetilde{v}_{00}-\sqrt{2}\widetilde{v}_{20}-\gamma\left(\widetilde{u}_{00}-\sqrt{2}\widetilde{u}_{20}\right)  = &~ 3\frac{\tau_{i}c}{2a}\widetilde{v}_{10}\\
\left(\frac{\tau_{s}}{\tau_{i}}-\gamma\right)\left(\widetilde{u}_{00}-\sqrt{2}\widetilde{u}_{20}\right)  = &~ 3\frac{\tau_{i}c}{2a}\widetilde{v}_{10}\\
\frac{\tau_{s}c}{2a}  = &~ \pm\frac{\tau_{s}}{\tau_{i}}\sqrt{\frac{1}{3}\left(\frac{\tau_{i}}{\tau_{s}}\gamma-1\right)}\label{eq:intrinsic_speed}
\end{align}
As shown in Figure \ref{fig:sub_c}, this formula can be verified 
by simulation.

\subsection{\label{subsec:moving_sub} Stability of Moving Bumps on 2D CANNs with SFA}

Using a method similar to that used in Appendix 
\ref{subsec:moving_std}, we define 
\begin{align}
K\left(\widetilde{u}_{00},\ldots,B|\widetilde{u}_{k_{0}k_{1}}\right) \equiv &~ -\widetilde{u}_{k_{0}k_{1}}+\frac{1}{B}\sum_{n_{0}m_{0}n_{1}m_{1}}C_{n_{0}m_{0}}^{k_{0}}C_{n_{1}m_{1}}^{k_{1}}\widetilde{u}_{n_{0}n_{1}}\widetilde{u}_{m_{0}m_{1}}-\widetilde{v}_{k_{0}k_{1}}\nonumber \\
   &~ -\frac{\tau_{s}c_{0}}{2a}\left(\sqrt{k_{0}}\widetilde{u}_{k_{0}-1,k_{1}}-\sqrt{k_{0}+1}\widetilde{u}_{k_{0}+1,k_{1}}\right)\nonumber \\
   &~ -\frac{\tau_{s}c_{1}}{2a}\left(\sqrt{k_{1}}\widetilde{u}_{k_{0},k_{1}-1}-\sqrt{k_{1}+1}\widetilde{u}_{k_{0},k_{1}+1}\right)\\
K\left(\widetilde{u}_{00},\ldots,B|\widetilde{v}_{k_{0}k_{1}}\right)  \equiv &~ \frac{\tau_s}{\tau_i} \biggl[-\widetilde{v}_{k_{0}k_{1}}+\gamma\widetilde{u}_{k_{0}k_{1}}\biggr.\nonumber \\
 &~-\frac{\tau_{i}c_{0}}{2a}\left(\sqrt{k_{0}}\widetilde{v}_{k_{0}-1,k_{1}}-\sqrt{k_{0}+1}\widetilde{v}_{k_{0}+1,k_{1}}\right)\nonumber \\
   &~ \biggl.-\frac{\tau_{i}c_{1}}{2a}\left(\sqrt{k_{1}}\widetilde{v}_{k_{0},k_{1}-1}-\sqrt{k_{1}+1}\widetilde{v}_{k_{0},k_{1}+1}\right)\biggr]\\
K\left(\widetilde{u}_{00},\ldots,B|B\right)  \equiv &~ \frac{\tau_s}{\tau_B} \biggl(-B\left(t\right)+1+\frac{1}{32\pi a^{2}}\widetilde{k}\sum_{k_{0}k_{1}}\widetilde{u}_{k_{0}k_{1}}^{2}\biggr).
\end{align}
Then, the matrix concerning the stability issue is 
\begin{equation}
\mathscr{A}_{\rm SFA}^{\rm moving}
 \equiv 
\left(\begin{array}{ccccc}
\frac{\partial K\left(\widetilde{u}_{00},\ldots,B|\widetilde{u}_{00}\right)}{\partial\widetilde{u}_{00}} & \cdots & \frac{\partial K\left(\widetilde{u}_{00},\ldots,B|\widetilde{u}_{00}\right)}{\partial\widetilde{v}_{00}} & \cdots & \frac{\partial K\left(\widetilde{u}_{00},\ldots,B|\widetilde{u}_{00}\right)}{\partial B}\\
\vdots & \ddots & \vdots & \ddots & \vdots\\
\frac{\partial K\left(\widetilde{u}_{00},\ldots,B|\widetilde{v}_{00}\right)}{\partial\widetilde{u}_{00}} & \cdots & \frac{\partial K\left(\widetilde{u}_{00},\ldots,B|\widetilde{v}_{00}\right)}{\partial\widetilde{v}_{00}} & \cdots & \frac{\partial K\left(\widetilde{u}_{00},\ldots,B|\widetilde{v}_{00}\right)}{\partial B}\\
\vdots & \ddots & \vdots & \ddots & \vdots\\
\frac{\partial K\left(\widetilde{u}_{00},\ldots,B|B\right)}{\partial\widetilde{u}_{00}} & \cdots & \frac{\partial K\left(\widetilde{u}_{00},\ldots,B|B\right)}{\partial\widetilde{v}_{00}} & \cdots & \frac{\partial K\left(\widetilde{u}_{00},\ldots,B|B\right)}{\partial B}
\end{array}\right).
\end{equation}
Similarly, the dynamics of distortions is given by
\begin{equation}
\tau_s
\frac{d}{dt}
\left(\begin{array}{c}
\delta \widetilde u_{00}\left(t\right) \\
\vdots \\
\delta \widetilde v_{00}\left(t\right) \\
\vdots \\
\delta B\left(t\right) \\
\end{array}\right)
=\mathscr{A}_{\rm SFA}^{\rm moving}
\left(\begin{array}{c}
\delta \widetilde u_{00}\left(t\right) \\
\vdots \\
\delta \widetilde v_{00}\left(t\right) \\
\vdots \\
\delta B\left(t\right) \\
\end{array}\right),
\end{equation}
Here $\delta \widetilde u_{k_0k_0}\left(t\right)$, $\delta \widetilde v_{k_0k_0}\left(t\right)$ and $\delta B\left(t\right)$ are defined by
\begin{align}
\widetilde u_{k_0k_1}\left(t\right) = &~ \widetilde u_{k_0k_1}^* + \delta \widetilde u_{k_0k_1} \left(t\right), \\
\widetilde v_{k_0k_1}\left(t\right) = &~ \widetilde v_{k_0k_1}^* + \delta \widetilde v_{k_0k_1} \left(t\right), \\
\widetilde B\left(t\right) = &~ B^* + \delta B \left(t\right),
\end{align}
where $\widetilde u_{k_0k_1}^*$, $\widetilde v_{k_0k_1}^*$ and $B^*$ are 
the fixed point moving solution to the system.
By calculating eigenvalues of $\mathscr{A}_{\rm SFA}^{\rm moving}$, stability of the fixed point solution can be determined.

\end{document}